\documentclass[a4paper,12pt]{article}
\usepackage[english]{babel}
\usepackage{a4wide}
\usepackage{amsmath,amsfonts,amssymb,amsthm,xcolor}
\usepackage{graphicx}
\usepackage[colorlinks=true, allcolors=blue]{hyperref}
\usepackage[toc, page]{appendix}
\usepackage{comment}
\usepackage{tikz}
\usetikzlibrary{arrows, shadows, automata, positioning, decorations.markings, calc}

\definecolor{purple}{rgb}{.765,.619,1}

\newtheorem{theorem}{Theorem}[section]
\newtheorem{proposition}{Proposition}[section]
\newtheorem{lemma}{Lemma}[section]
\newtheorem{corollary}{Corollary}[section]
\newtheorem{definition}{Definition}[section]
\newtheorem{assumption}{Assumption}[section]

\title{\bf On the Estimation of the Time-Dependent Transmission Rate in Epidemiological Models}
\author{Jorge P. Zubelli\thanks{Mathematics Department, Khalifa University, Abu Dhabi, UAE, \href{mailto:jorge.zubelli@ku.ac.ae}{\tt jorge.zubelli@ku.ac.ae}} \thanks{ADIA Lab, Abu Dhabi, UAE.}, \, Jennifer Loria\thanks{CIMPA -E. de Matemática, U. de Costa Rica, Costa Rica, \href{mailto:jenniferdelosange.loria@ucr.ac.cr}{\tt jenniferdelosange.loria@ucr.ac.cr}},\, and Vinicius V. L. Albani\thanks{Federal University of Santa Catarina, \,   
88.040-900 Florianopolis, Brazil, \href{mailto:v.albani@ufsc.br}{\tt v.albani@ufsc.br}}}
\date{\today}

\begin{document}

\maketitle

\begin{abstract}
The COVID-19 pandemic highlighted the need to improve the modeling, estimation, and prediction of how infectious diseases spread. SEIR-like models have been particularly successful in providing accurate short-term predictions.

This study fills a notable literature gap by exploring the following question: Is it possible to incorporate a nonparametric susceptible-exposed-infected-removed (SEIR) COVID-19 model into the inverse-problem regularization framework when the transmission coefficient varies over time?

Our positive response considers varying degrees of disease severity, vaccination, and other time-dependent parameters. In addition, we demonstrate the continuity, differentiability, and injectivity of the operator that link the transmission parameter to the observed infection numbers.

By employing Tikhonov-type regularization to the corresponding inverse problem, we establish the existence and stability of regularized solutions. Numerical examples using both synthetic and real data illustrate the model's estimation accuracy and its ability to fit the data effectively.
\vspace{6pt}

\end{abstract}

\section{Introduction}

In December 2019, the first cases of the severe acute respiratory syndrome coronavirus 2 (SARS-CoV-2), that caused the coronavirus disease 2019 (COVID-19), were detected in China. After a rapid spread across the world, the World Health Organization (WHO) characterized this outbreak as a pandemic. More than three years later, in May 2023, WHO considered that COVID-19 could no longer be classified as a Public Health Emergency of International Concern (PHEIC) \cite{WHO}. To illustrate the impact of the pandemic, by 13 April 2024, SARS-CoV-2 caused more than seven million deaths and infected more than 700 million people around the world \cite{worldometer}.

Modeling the evolution of pandemics at different levels became one of the most important research topics at the time. A large amount of models to describe, estimate, and predict the SARS-CoV-2 dynamics were proposed. Different approaches were considered, such as agent and network-based models \cite{kerr2021,achterberg2020}, susceptible-infected-removed-type (SIR-like) models \cite{albani2022,Albani2020,beretta2021,bertozzi2020,campos2021,gatto2020}, models based on partial differential equations \cite{guglielmi2022}, approaches based on statistical modeling and neural networks \cite{stewart2022,namasudra2021}, or even multiscale approaches \cite{B1,bellomo2}. Forecasting accurately the spread dynamics of infectious diseases is useful in helping public authorities design appropriate mitigation or contention measures. Lockdowns, for example, if implemented for a long time, can have a negative socio-economic impact \cite{albani2022role,albani2022c}. Forecast performance can be linked to model parsimony \cite{bertozzi2020}. Fancy models, with too many degrees of freedom, can overfit data and produce inaccurate forecasts. This means that models must be well designed to account for a good representation of the dynamics and forecast accuracy.

The transmission parameter in SIR-like or susceptible-exposed-infected-removed-like (SEIR-like) models depends on the average of contacts an individual has in a period and on the probability of contact with an infected individual results in transmission. Such quantities, in turn, depend on mobility and the in-host pathogen dynamics \cite{B1,bellomo2,keeling2008} that change with time. Thus, a time-dependent transmission parameter is a natural hypothesis. Some works explored this assumption and estimated the evolution of this parameter from COVID-19 data providing empirical evidence \cite{albani2022,albani2021covid,Albani2021,Albani2020,athayde2022,somersalo2020,calvetti2020}.

Based on this premise, we analyze the direct and inverse problems associated with an SEIR-like model with a time-dependent transmission parameter. More precisely, after appropriately defining the parameter-to-solution map that associates the transmission parameter to the number of daily infections, regularity properties, such as well-posedness, continuity, and differentiability, are stated. Applying Tikhonov-type regularization to the problem of estimating the transmission parameter from reported daily infections, sufficient conditions for the existence, stability, and convergence of solutions are provided, using variational methods \cite{scherzer2009}. Thus, the main contribution of the present article is to provide a rigorous analysis of the problem of estimating time-dependent transmission parameters that arise in the transmission of infectious diseases.
It should be mentioned that the proposed SEIR-like model is designed to describe the spread of the SARS-CoV-2 virus. Thus, it accounts for different severity levels. Moreover, most of the model parameters can be obtained in the literature, however, the transmission rate is unknown and must be estimated from the observed incidence.

This article is organized as follows: Section \ref{sec:model} presents the proposed SEIR-like model with its parameters and compartments. In section \ref{sec_pr_dir}, it is shown that the direct problem has a unique and non-negative solution in the set $I=[0,T]$, for any $T > 0$, including the case $T = \infty$. The parameter-to-solution is defined and its regularity properties are stated. 
 In Section~\ref{sec:inverse}, the inverse problem of estimating the transmission parameter is analyzed. To illustrate the theoretical results, Section~\ref{sec:numerics} presents a numerical example with real data.
 

\section{The Proposed Model}\label{sec:model}

\begin{figure}[h]
\begin{tikzpicture}[->,>=stealth',shorten >=1pt,auto,node distance=3.5cm, on grid,
semithick, every state/.style={fill=purple,draw=none,scale=1.5,circular drop shadow,text=black}]
\node[state, xshift= 4 cm, fill=white]               (S) at (0,0)	            {$S$};
\node[state, fill= gray!60]		                   (E) [right=of S]    {$E$};
\node[state, yshift=0.5 cm, fill=orange!30]		                   (V) [below=of S]    {$V$};
\node[state, fill=blue!30]		                   (IM) [above right =of E]    {$I_M$};
\node[state,fill=blue!45]		                   (IS) [right =of IM]    {$I_S$};
\node[state, fill=blue!60]		                   (IC) [right =of IS]    {$I_C$};
\node[state, fill=yellow!80]		                   (IA) [below right =of E]    {$I_A$};
\node[state, fill=green!70]		                   (R) [right =of IA]    {$R$};
\node[state, fill=black, text=white]		                   (D) [right =of R]    {$D$};

\path (S) edge	node         { }(V)
(S) edge	node   { } (E)
(E) edge	node   { } (IM)
(E) edge	node   { } (IA)
(IM) edge	node   { } (IS)
(IS) edge	node   { } (IC)
(IC) edge	node   { } (D)
(IA) edge	node   { } (R)
(IM) edge	node   { } (R)
(IS) edge	node   { } (R)
(IC) edge	node   { } (R);
\end{tikzpicture}
\caption{Schematic representation of the epidemiological model of Eqs. (1)-(9).}
\label{Fig_modelo}
\end{figure}
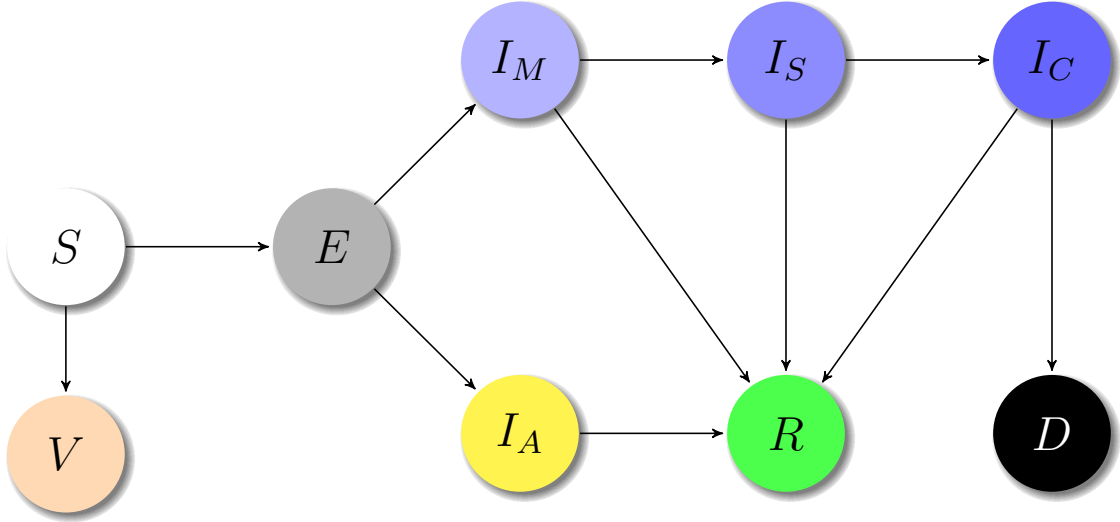

The SEIR-Like model describes the dynamics of the spread of the SARS-CoV-2 virus in a susceptible population. It accounts for nine compartments, namely, susceptible ($S$), exposed ($E$), vaccinated ($V$), asymptomatic infectious ($I_A$), mildly infectious ($I_M$), severely infectious ($I_S$), critically infectious ($I_C$), recovered ($R$), and deceased ($D$). A summary of the model compartments description can be found in Table~\ref{tab:compartments} and Figure~\ref{Fig_modelo} shows the schematic representation of the interaction between the model compartments. 

The system of ordinary differential equations (ODE) that determines the time evolution of the model compartments is the following:
\begin{align}\label{model_1}
    \frac{dS}{dt} = -S(\beta_A I_A + \beta_M I_M + \beta_S I_S + \beta_C I_C) - \nu S,\\
    \frac{dV}{dt} = \nu S,\\
    \frac{dE}{dt} = S(\beta_A I_A + \beta_M I_M + \beta_S I_S + \beta_C I_C) - \sigma E,\\
    \frac{dI_A}{dt} = (1-p)\sigma E - \gamma_{R,A}I_A,\\
    \frac{dI_M}{dt} = p\sigma E - (\gamma_{R,M} + \alpha_S)I_M,\\
    \frac{dI_S}{dt} = \alpha_S I_M - (\gamma_{R,S} + \alpha_C) I_S,\\
    \frac{dI_C}{dt} = \alpha_C I_S - (\gamma_{R,C} + \delta_D) I_C,\\
    \frac{dR}{dt} =\gamma_{R,A} I_A + \gamma_{R,M} I_M + \gamma_{R,S} I_S + \gamma_{R,C} I_C,\\
    \frac{dD}{dt} = \delta_D I_S. \label{model_2}
\end{align}

\begin{table}[!htb]
\centering
\begin{tabular}{|c|p{13cm}|}
\hline
Symbol & Description\\
\hline
$S(t)$ & Susceptible individuals: not infected yet.\\
$E(t)$ & Exposed: i.e., infected but not infectious individuals.\\
$V(t)$ & Vaccinated: individuals who received a vaccine and are immunized against infection.\\
$I_A(t)$ & Asymptomatic infectious: infectious with no symptoms.\\
$I_M(t)$ & Mildly infectious: symptomatic infectious with mild symptoms.\\
$I_S(t)$ & Severely infectious: symptomatic infectious, admitted to a regular hospital bed.\\
$I_C(t)$ & Critically Infectious: symptomatic infectious, admitted to an intensive care unit (ICU).\\
$R(t)$ & Recovered: individuals that recovered from infection and are immune.\\
$D(t)$ & Deceased: individuals who died from the disease.\\
\hline
\end{tabular}
\caption{Model compartments description.}
\label{tab:compartments}
\end{table}

\begin{table}[!htb]
\begin{center}
\resizebox{\textwidth}{!}{
\begin{tabular}{|c|p{8cm}|c|c|}
\hline
Symbol &  Description & Value & Reference\\
\hline
$\beta_A$ & Transmission rate for asymptomatic infectious individuals. & $0.58\beta_M$ & \cite{byambasuren2020}\\
$\beta_M$ & Transmission rate for mildly infectious individuals. & Estimated & -\\
$\beta_S$ & Transmission rate for severely infectious individuals. & $0.1\beta_M$ & \cite{Albani2020} \\
$\beta_C$ & Transmission rate for critically infectious individuals. & $0.01\beta_M$ & \cite{Albani2020}\\
$1/\gamma_{R,A}$ & Recovery rate of asymptomatic infectious individuals. & $14$ days & \cite{who19}\\
$1/\gamma_{R,M}$ & Recovery rate of mildly infectious individuals. & $14$ days & \cite{who19}\\
$1/\gamma_{R,S}$ & Recovery rate of severely infectious individuals. & $12$ days & \cite{guan2020}\\
$1/\gamma_{R,C}$ & Recovery rate of critically infectious individuals. & $9$ days & \cite{grasselli2020}\\
$\alpha_S$ & Hospitalization rate. & $\dfrac{\hat{H}(t)}{\hat{I}(t-1)}$ & \cite{lauer2020}\\
 $\alpha_C$ & ICU admission rate. & $0.4$ & \cite{abate2020}\\
$1/\sigma$ & Meantime from contagion to become infectious. & $5.1$ days & \cite{lauer2020}\\
$\nu$ & Vaccination rate. & $0$ & -\\
$p$ & Proportion of exposed individuals becoming mildly infectious. & $0.83$ & \cite{byambasuren2020}\\
$\delta_D$ & Death from infection rate. & $\dfrac{\hat{D}(t)}{\alpha_C(t)\hat{H}(t-11)}$ & \cite{huang2020,grasselli2020}\\
\hline
\end{tabular}}
\caption{Parameters representation of the epidemiological model of Eqs. (1)-(9).}
\label{Tab_Par}
\end{center}
\end{table}

The transmission rates for asymptomatic, mildly, severely, and critically infective individuals are indicated by $\beta_A$, $\beta_M$, $\beta_S$, and $\beta_C$, respectively. The vaccination rate is $\nu$, which is the product of the daily vaccination rate of susceptible individuals and the effectiveness against infection of the vaccine used. The mean time from contagion to become infectious is the inverse of the parameter $\sigma$. The recovery rate of mildly, severely, critical and asymptomatic infective individuals is indicated by $\gamma_{R,M}$, $\gamma_{R,S}$, $\gamma_{R,C}$, and $\gamma_{R,A}$, respectively. The rates of hospitalization and ICU admission are denoted by $\alpha_S$ and $\alpha_C$, respectively. According to the World Health Organization, only people in critical conditions generally die of COVID-19, therefore the corresponding death rate is $\delta_D$ \cite{who19}. Table \ref{Tab_Par} summarizes the description of the model parameters. The parameters $\beta_A$, $\beta_M$, $\beta_S$, $\beta_C$, $\gamma_{R,M}$, $\gamma_{R,S}$, $\gamma_{R,C}$, $\gamma_{R,A}$, $\alpha_S$, $\alpha_C$, and $\delta_D$ depend on time.

In general, the parameters are $\beta_A$, $\beta_M$, $\beta_S$, and $\beta_C$, as well as the initial number of mildly and asymptomatic infective individuals, are unknown and estimated from, for example, the daily numbers of infections, deaths, hospital and ICU admissions. However, in this article only the estimation of the transmission rates will be considered. For simplicity and to reduce the number of unknowns, it is assumed that all the transmission parameters are parameterized by the transmission associated with mild infectious individuals:
\begin{equation}\label{eq_betas}
\beta_S= a\beta_M \text{, }\beta_C= b\beta_M \text{, and  } \beta_A= c\beta_M, 
\end{equation}
with $a=0.1$, $b=0.01$, and $c=0.58$. This means that the infection rate of hospitalized, in ICU and asymptomatic individuals are 10\%, 1\%, and 58\%, respectively, of the transmission rate of those in the mildly infective compartment \cite{Albani2020, byambasuren2020}. The mean time between infection and becoming infective is set to 5.1 days \cite{lauer2020}. The proportion of exposed individuals becoming mildly infective is $p$, which is set to 0.83 \cite{byambasuren2020}. All recovery rates for mildly, severely, critically infective, and asymptomatic individuals are set at 1/14 days$^{-1}$. 
The admission rate to the ICU is set as $\alpha_C=0.4$ \cite{abate2020}. The hospitalization and death rates are  defined, respectively, as the following proportions:
\begin{equation}
\alpha_S(t)= \frac{\hat{H}(t)}{\hat{I}(t-1)} \text{ and } \delta_D(t)=\frac{\hat{D}(t)}{\alpha_C(t)\hat{H}(t-11)}
\end{equation}
where $\hat{I}$, $\hat{H}$, and $\hat{D}$ represent the time series of the seven-day moving average of daily reported infections, hospitalizations, and deaths, respectively. Here, a delay of one day from the onset to hospitalization \cite{lauer2020} days and an 11-day delay from hospitalization to death \cite{huang2020,grasselli2020} are considered.
Notice that, all the model parameters are non-negative.


\subsection{Model Well-Posedness}
To analyze the direct problem, it is necessary to state the well-posedness of the proposed SEIR-like model. The following assumption will be necessary in the analysis that follows.

\begin{assumption}\label{hip_solutions}
All the parameters of the SEIR-like model in Eqs~\eqref{model_1}--\eqref{model_2} are continuous and strictly positive. 
\end{assumption}

As mentioned before, all the model parameters are positive. It is also expect that they depend continuously in time, as variations in the disease progression and the virus spread tend to be stable and well-behaved for small periods. However, transmission could jump due to strong contention measures or the sudden emergence of a highly infectious variant \cite{albani2022,Albani2024}. In principle, such a function with jumps can be approximated by a continuous function.

For simplicity, we denote the dependent variables of the system in Eqs~\ref{model_1}--\ref{model_2} as follows:
\[x_1=S, x_2=V,\cdots,x_9=D.\]


Let the initial condition $\boldsymbol{x}_0=(x_{1}(0),...,x_{9}(0))$ be nonnegative, i.e. $x_{i} \geq 0$ ($i=1,...,9$). Moreover, assume that the initial susceptible and exposed populations are positive, that is, $x_{1},x_{3} > 0$. In what follows,  $\mathbb{R}^n$ is endowed with the $2$-norm.

\begin{definition} 
\label{def_f} Let $\boldsymbol{f}:D(\boldsymbol{f})\subset \mathbb{R} \times \mathbb{R}^9\rightarrow \mathbb{R}^9$, with $D(\boldsymbol{f})=(-T_0,T_0) \times B_{\delta_0}(\textbf{\textit{x}}_0)$, $T_0 > T$, and $\delta_0 = 2 \| \textit{\textbf{x}}_0\|_1 +1$, be the function defined by the right-hand side (RHS) of the system in Eqs~\eqref{model_1}--\eqref{model_2}.
%
\end{definition}
Notice that $B_{\delta_0}(\textbf{\textit{x}}_0)$ is the open ball in $\mathbb{R}^9$, centered at $\textbf{\textit{x}}_0$ with radius $\delta_0$. 
The initial value problem associated with the system in Eqs~\eqref{model_1}--\eqref{model_2} can be written as:
\begin{equation}
    \label{re_edo}
    \left \{ \begin{matrix} \boldsymbol{x}^\prime=\boldsymbol{f}(t,\boldsymbol{x}), \quad t\in I,\\ 
\boldsymbol{x}(0)=\boldsymbol{x}_0. \end{matrix} \right.
\end{equation}


In what follows, it is firstly shown that any solution for the ODE problem in Eq.~\ref{re_edo} must be non-negative. After, existence and uniqueness of solution is stated.


\begin{proposition} \label{pr2}
Let $\boldsymbol{u}$ be a continuous solution of the ODE problem in Eq.~\eqref{re_edo} and defined in the interval $J=[0,T_{0})$, with $0<T<T_{0}\leq \infty$.  
Let also Assumption~\ref{hip_solutions} hold. Then, $u_{1}(t) > 0$ , $\forall t \in J$.
\end{proposition}


 

\begin{proof}
 Suppose the statement is false Since $u_{1}(0) > 0$, there exists an $s>0$ such that $u_{1}(s)=0$ and $u_1(t)>0$ for all $t\in [0,s)$ and $u_{1}^\prime(t)/u_{1}(t)$ is well defined for $t\in [0,s)$. For any $t\in[0,s)$, define the function 
 $$G(t)= -(\beta_A(t) u_{4}(t) + \beta_M(t) u_{5}(t)+ \beta_S(t) u_{6}(t) + \beta_C(t) u_{7}(t) + \nu(t)).$$
Since all the parameters and the and $u_{i}$'s are continuous, $G$ is continuous. For any $t\in [0,s)$, it follows that
$$\displaystyle \frac{ u_{1}^\prime(t)}{u_{1}(t)}= 
G(t), \quad \mbox{and thus,} \quad
u_{1}(t)= u_{1}(0) \exp \left(\int _{0}^{t} G(y) dy\right).$$
By the definition of $G$, the limit
$\lim _{t \rightarrow s^-}\int _{0}^{t} G(y) dy$ exists and is finite. Thus, on one hand,
$\lim _{t \rightarrow s^-}  u_{1}(t) = \lim _{t \rightarrow s^-} u_1(0) \exp\left(\int _{0}^{t} G(y) dy\right)>0.$ 
On the other hand, by the continuity of $u_1$ and the hypothesis $u_1(s)=0$, and we have a contradiction.
%
\end{proof}
%

\begin{proposition}\label{pr3}
Let the hypotheses of Proposition~\ref{pr2} hold. 
Let also $i$ be in $\{ 4,5,6,7 \}$. If $s$ represents the first zero of $u_{3}$ or the first infinity value in the case of $u_3$ be strictly positive, then if there exists an $s_0 < s$ such that $u_i(s_0)> 0$, then $u_{i}(t) > 0$ for all $t \in [s_0,s)$.






\end{proposition}

\begin{proof}
 Only the case $u_4 > 0$ is shown; the others are similar. 
 Suppose, by contradiction, that $u_4$ is not strictly positive in $[s_0,s)$. Since $u_4$ is continuous and $u_4(s_0)>0$, there exists $t_0\in(s_0,s)$ such that $u_4(t_0)=0$ and $u_4(t)>0$ for $t\in [s_0,t_0)$. 
Notice that
\[u_4^\prime(t_0)= (1-p)\sigma u_{3}(t_0) - \gamma_{R,A}(t_0) u_{4}(t_0) = (1-p)\sigma u_{3}(t_0)>0,\]
because $(1-p)\sigma >0$ and $u_3(t_0) > 0$.

Since $u^\prime_4$ is continuous, there exists an $\varepsilon >0$ such that $u_4^\prime(t) > 0$ for $t\in(t_0-\varepsilon, t_0 +\varepsilon) \subset [s_0,s)$ and $u_4$ is strictly increasing in $(t_0-\varepsilon, t_0 +\varepsilon)$. This implies $0<u_4(t_0 - \varepsilon) < u_4(t_0) = 0$, which is a contradiction.
\end{proof}

The previous proposition implies that whenever $u_i$ ($i=4,5,6,7$) is positive, it will remain positive while $u_3$ is also positive. For $i=4,5,6,7$, $u_i$ represents the infectious compartments and $u_3$ represents the exposed compartment. 
In other words, while there are exposed individuals, there will also be infectious individuals.

\begin{proposition}\label{lem3}
Let the hypotheses of Proposition~\ref{pr2} hold.
If $i=4,5,6,7$, then,
\begin{enumerate}
\item[(i)] If $u_i (0)=0$, with $i=4,5$, then there exists $\varepsilon > 0$ such that $u_i(t) > 0$, $\forall t \in (0, \varepsilon)$.
\item[(ii)] If $u_i (0) =0$, with $i=6,7$, then there exists $\varepsilon > 0$ such that $u_i(t) \geq 0$, $\forall t \in (0, \varepsilon)$.
\end{enumerate}
\end{proposition}

\begin{proof}
Assume that $u_4(0)=0$. Notice that $u_4^\prime(0)= (1-p)\sigma u_{3}(0) - \gamma_{R,A}(0) u_{4}(0) = (1-p)\sigma u_{3}(0)>0$. Thus, by the continuity of $u^\prime_4$, there exists $\varepsilon >0$ such that $u^\prime_4 > 0$ in $[0, \varepsilon)$. Thus, $u_4$ is strictly increasing in $(0, \varepsilon]$ and $0=u_4(0) < u_4(t)$, for $t \in (0, \varepsilon]$. The proof for $u_5$ is analogous.

Suppose that $u_6(0)=0$. Since $u_3(0)>0$, there exists $\varepsilon > 0$ such that $u_3 >0$ in $[0,\varepsilon]$. Suppose, by contradiction, that there exists $t_0\in (0,\varepsilon]$ such that $u_6(t_0)<0$. Then $u_6(t) \leq 0$, $\forall t \leq t_0$, otherwise since $u_3(\varepsilon) >0$ by Prop.~\ref{pr3} $u_6(t_0)>0$.
By Assumption \ref{hip_solutions}, the previous item (take the smaller $\varepsilon$, if necessary), and Prop.~\ref{pr3},  it follows that, for $t \in[0,t_0]$,
$$u_6^\prime(t) = \alpha_S(t) u_{5}(t) - (\gamma_{R,S}(t) + \alpha_C(t)) u_{6}(t) \geq 0.$$  
Thus, $u_6$ is non-decreasing in $[0,t_0]$ and $0=u_6(0) \leq u_6(t_0) < 0 $, which is a contradiction. Therefore, $u_6 \geq 0$ in $(0, \varepsilon)$ and the proof for $u_7$ is analogous. 
\end{proof}

The following corollary is an immediate consequence of Propositions~\ref{pr3}--\ref{lem3}.
\begin{corollary} \label{cor1}
Let the hypotheses of Proposition~\ref{pr2} hold. 
If $i=4,5,6,7$, then there exists $\varepsilon > 0$ such that $u_i(t) \geq 0$, for any $t \in (0, \varepsilon)$ . Furthermore, if $u_3$ has a first zero at $s$, then, $\varepsilon = s$.
\end{corollary}
%
\begin{proposition}\label{pr4}
Let the hypotheses of Proposition~\ref{pr2} hold. Then $u_{3}(t) > 0$, $\forall t \in J$.
\end{proposition}
\begin{proof}
Suppose, by contradiction, that $u_3$ is not strictly positive in $J$. Since $u_3(0)>0$ and $u_3$ is continuous, there exists $s>0$ in $J$ such that $u_3(s)=0$ and $u_3(t)>0$ for any $t\in[0,s)$. Moreover,
$u_{3}^\prime (s) 
=u_{1}(s)\beta_M(s) (c u_{4}(s)+  u_{5}(s)+ a u_{6}(s) + b  u_{7}(s))$. By Prop.~\ref{pr2} $u_1(s) > 0$ and by Assumption \ref{hip_solutions} $\beta_M(s)>0$. By Corollary \ref{cor1} $c u_{4}(s)+  u_{5}(s)+ a u_{6}(s) + b  u_{7}(s) \geq 0$ . Thus, $u_{3}^\prime (s) \geq 0$.

If $u_{3}^\prime (s) > 0$, as in the proof of Prop.~\ref{pr3}, there exists $\varepsilon >0$ such that $0<u_3(s - \varepsilon) < u_3(s) = 0$, which is a contradiction.

If $u_{3}^\prime(s)=0$, it follows that $c u_{4}(s)+  u_{5}(s)+ a u_{6}(s) + b  u_{7}(s) =0$, which implies that $u_{4}(s)= u_{5}(s)= u_{6}(s) = u_{7}(s)=0$, since theses functions are non-negative and the parameters are strictly positive. Define the following linear ODE system with $u_1$ given:
\begin{align}
    \frac{dy_{1}}{dt} = -u_{1}(\beta_A y_{2} + \beta_M y_{3}+ \beta_S y_{4} + \beta_C y_{5}) - \sigma y_{1}\label{eq3}\\
    \frac{dy_{2}}{dt} = (1-p)\sigma y_{1} - \gamma_{R,A} y_{2}\\
    \frac{dy_{3}}{dt} = p\sigma y_{1} - (\gamma_{R,M} + \alpha_S) y_{3}\\
    \frac{dy_{4}}{dt} = \alpha_S y_{3} - (\gamma_{R,S} + \alpha_C) y_{4}\\
    \frac{dy_{5}}{dt} = \alpha_C y_{4} - (\gamma_{R,C} + \delta_D) y_{5}\label{eq4}
\end{align}
%

%
If the initial condition for the system in Eqs.~\eqref{eq3}--\eqref{eq4} is $\boldsymbol{y}(s)=\boldsymbol{0}$, then, $\boldsymbol{y}\equiv \boldsymbol{0}$ is a solution in $J$. On the other hand, $\boldsymbol{y}=(u_3,u_4,u_5,u_6,u_7)$ is also a solution for the same problem in $J$. To see this, compare Eqs.~\eqref{eq3}--\eqref{eq4} with the system in Eqs.~\eqref{model_1}--\eqref{model_2} and use the fact that $u_{4}(s)= u_{5}(s)= u_{6}(s) = u_{7}(s)=0$. 
%
By the uniqueness of solutions \cite[Corollary~ 1.11]{sotomayor2011}, it follows that $u_{4}= u_{5}= u_{6}= u_{7}=0$ in $J$, which is a contradiction. 
Therefore, $u_3>0$ in $J$. 
\end{proof} 

\begin{proposition} \label{pr5}
Let the hypotheses of Proposition~\ref{pr2} hold. 
Then, $u_i > 0$, with $i=1,3$, and $u_i \geq 0$, with $i=2,4,5,6,7,8,9$, in $J$.
\end{proposition}
\begin{proof}
The propositions~\ref{pr2} and \ref{pr4} imply that $u_1>0$ and $u_3>0$ in $J$. Proposition \ref{pr3} and Lemma \ref{lem3} imply that $u_i \geq 0$ in $J$, with $i =4,5,6,7$. 

Since $u_2^\prime= \nu u_{1} \geq 0$ in $J$, it follows that $u_2$ is increasing. Therefore, $0 \leq u_2(0) \leq u_2(t)$, $\forall t \in J$. With an analogous argument, it follows that $u_8 \geq 0$ and $u_9 \geq 0$ in $J$.
%
%
\end{proof}

Defining $N = \sum_{i=1} ^9 u_i$, under the hypotheses of Proposition~\ref{pr2}, it follows that $N$ is constant since $N^\prime=\displaystyle \sum_{i=1} ^9  u_i^\prime=0$.

%

\begin{proposition}\label{pr6}
Let Assumption \ref{hip_solutions} hold. Then, the ODE system in Eq.~\eqref{re_edo} has a unique solution in the maximal interval.
\end{proposition}

\begin{proof}
 The function $\boldsymbol{f}$ is continuous since all the parameters are continuous and the dependence on the $x_i$'s is polynomial. Moreover, $\boldsymbol{f}$ is locally Lipschitz continuous in the second argument, uniformly with respect to $t$. Indeed, let $V_0 \subset D(\boldsymbol{f})$ be a compact set and $(t,\boldsymbol{x}),(t,\boldsymbol{y})$ be two elements of $V_0$. Define $M=\max_{(t,\boldsymbol{x}) \in V_0} \|\boldsymbol{x}\|_2$ 
 and $K$ as the supremum value of all model parameters defining $\boldsymbol{f}$ in $J=[0,T_0)$. 

Setting $L'= \max (8M(K+1), 4K)$, it follows that $\|\boldsymbol f(t,\boldsymbol x) -\boldsymbol f(t,\boldsymbol y)\|_2 \leq 9 L' \|\boldsymbol x-\boldsymbol y\|_2$ for any $(t,\boldsymbol{x}),(t,\boldsymbol{y})$ in $V_0$. Therefore, 
%
$$\sup _{(t,x) \neq (t,y) \in V_0} \frac{\|\boldsymbol f(t,\boldsymbol x) -\boldsymbol f(t,\boldsymbol y)\|_2}{\|\boldsymbol x-\boldsymbol y\|_2} <\infty \text{, for } x\neq y.$$

By the Picard–Lindelöf Theorem \cite{teschl2012}, there exists a unique local solution $\overline{u} \in C^1 (\tilde J)$  of the ODE system in Eq.~\eqref{re_edo}, where $\tilde J$ is some interval around $0$. Moreover, by \cite[Theorem~2.13]{teschl2012}, the interval $\tilde J$ is maximal.
\end{proof}
In what follows, we assume that $T_0$ is the right edge of the interval $\tilde J$.
\begin{proposition}\label{pr7}
The interval $I=[0,T]$ is a subset of $\tilde J$.
\end{proposition}

\begin{proof}
Notice that $\tilde J =(t_-, t_+)$. Thus, $t_+ \geq T_0$ must hold. Suppose, by contradiction, that $t_+ < T_0$. 
Take $\delta_1\in (2\|\boldsymbol{x}_0\|_1,2\|\textit{\textbf{x}}_0\|_1+1)$ and $C=\overline{B}_{\delta_1}(\textit{\textbf{x}}_0)$, the closed ball centered at $\boldsymbol{x}_0$ with radius $\delta_1$. Let $\{t_n\}_{n \in \mathbb{N}}$ be a sequence in $(0,t_+)$ with $\lim_{n \to \infty} t_n =t_+$. 
Let $\overline{\boldsymbol{u}}$ be the maximal solution of the ODE system in Eq.~\eqref{re_edo} defined in $\tilde J$. Since $\overline{u}_i\geq 0$ ($i=1,\cdots,9)$ and $\overline{N}=\sum_{i=1}^9\overline{u}_i$ is constant, then $\|\overline{\boldsymbol{u}}(t_n)\|_1 = \overline{N} = \|\overline{\boldsymbol{u}}(0)\|_1=\|\boldsymbol{x}_0\|_1$. 
Thus, $\|\overline{\boldsymbol{u}}(t_n)-\textit{\textbf{x}}_0\|_1\leq 
2 \|\textit{\textbf{x}}_0\|_1 < \delta_1$ and 
$(t_n,\overline{\boldsymbol u}(t_n)) \in [0, t_+] \times C \subset D(\boldsymbol f)$ for all $n \in \mathbb{N}$.

By \cite[Corollary 2.15]{teschl2012}, there exists an extension of the solution to the interval $(t_- , t_+ + \varepsilon)$, for some $\varepsilon > 0$, which is a contradiction since $\tilde J=(t_-,t_+)$ is maximal. Therefore, $t_+\geq T_0$ and there is a unique solution $\overline{\boldsymbol u}$ defined in $I=[0,T]$, with $T\in (0,t_+)$.
\end{proof} 

A natural question is the existence of a unique nonnegative solution $\overline{\boldsymbol u}$ in $I= [0,\infty)$. If the parameters are also bounded in $I$, then the answer is positive.

\begin{corollary}
If all the parameters in the ODE system~\eqref{re_edo} are bounded in $[0,+ \infty)$, then there exists a unique non-negative solution in $[0,+\infty)$.
\end{corollary}

\begin{proof}
The proof is similar to the proof of Proposition~\ref{pr6}. 
%
To show that $[0,\infty) \subseteq \tilde J$, note that the compact set $C$ in the proof of Proposition \ref{pr7} does not depend on $t_+$ and apply \cite[Corollary 2.15]{teschl2012}.
\end{proof}

\section{The Parameter-to-Solution Map}\label{sec_pr_dir}
This section analyzes the direct problem and provides the well-posedness and regularity properties of the parameter-to-solution map. Consider the closed interval $I=[0,T]$ with $T >0$, the Sobolev space $X =H^{1}(I)$ with the norm $\Vert\varphi\Vert_{X}^2=\Vert\varphi\Vert_{L^2(I)}^2 + \Vert\varphi'\Vert_{L^2(I)}^2$, for any $\varphi \in X$ with weak derivative $\varphi^\prime$. Define also the vector Sobolev space $Y=H^{1}(I)^n$ with $n\geq 1$ and norm defined as
\[\displaystyle \Vert \boldsymbol\phi \Vert_{Y}^2=\Vert \phi _1 \Vert_{H^1(I)}^2 +...+ \Vert \phi_n \Vert_{H^1(I)}^2,\]
for any $\boldsymbol\phi \in Y$. For the problem under consideration, $n=9$ in the $Y$ definition. 

Let $\lambda \geq 0$ be fixed and define the parameter-to-solution map domain as
\[D_{\lambda}(\mathcal{J})=\{ \beta \in H^1 (I): \beta \geq \lambda \text{ almost everywhere in } I \}.\]
Since all the transmission parameters in the model are parameterized by $\beta_M$, for simplicity, we denote this parameter by $\beta$.

By \cite[Theorem~8.2]{brezis2011}, for every $\varphi \in H^1(I)$, there exists a continuous function $\overline{\varphi} \in H^1(I)$ with $\varphi=\overline{\varphi}$  almost everywhere (a. e.)  in $I$ which is called a continuous representative of $\varphi$. Thus, without loss of generality, from now on we assume that any $\beta \in D_{\lambda}(\mathcal{J})$ is continuous. 
Thus, $\beta(t) \geq \lambda$, $\forall  t \in I$. 

Before defining the parameter-to-solution operator $\mathcal{J}$ appropriately, we need some preliminary results.

\begin{proposition}
    For any $\lambda\geq 0$, the set $D_{\lambda}(\mathcal{J})$ has a non-empty interior and is closed and convex.
    \label{pr_in_Novacio}
\end{proposition}
\begin{proof}
By Sobolev inequality \cite[p.~212]{brezis2011}, there exists a constant $C>0$, depending only on $I$, such that $\Vert \varphi \Vert _{L^{\infty}(I)} \leq C \Vert \varphi \Vert _{H^{1}(I)}$, $\forall \varphi\in H^{1}(I)$. 
For $\lambda > 0$, define $g(t)=2 \lambda e^{t}$ in $[0,T]$ and $\varepsilon = \lambda/(3C)$. Thus $\lambda < 2 \lambda\leq 2\lambda e^t$, $\forall t \in [0,T]$, and $g \in D_\lambda(\mathcal{J})$. Denote the open ball in $H^{1}(I)$, centered in $g$ with radius $\varepsilon$, by $B=B_{\varepsilon}(g)$. 

Let $h$ be an arbitrary element of $B$. 
Suppose, by contradiction, that there exists $t_0 \in I$ such that $h(t_0) - \lambda<0$. Then, by the continuity of $h$, there exists an interval $I_0 \subset I$ such that $h(t)-\lambda< 0$ in $I_0$. Thus, for any $t \in I_0$, it follows that
\[|g(t)-h(t)| = g(t)-h(t) =2\lambda e^t -h(t) \geq \lambda.\]
Thus, $\lambda \leq \Vert g-h \Vert _{L^{\infty}(I)} \leq C  \Vert g-h \Vert _{H^{1}(I)}$ and then, $\Vert g-h \Vert _{H^{1}(I)}>\varepsilon$, which is a contradiction and $h \in D_{\lambda}(\mathcal{J})$. The case $\lambda=0$, follows similarly, just consider $g(t)=e^t$ and $\varepsilon= 1/(3C)$. Therefore, $D_{\lambda}(\mathcal{J})$ has a non-empty interior.

Let $\{\beta_n\}_{n\in \mathbb{N}}$ be a sequence in $D_{\lambda}(\mathcal{J})$ converging in norm to $\beta$ in $H^{1}(I)$. Since $\beta_n \in D_{\lambda}(\mathcal{J})$ for all $n$, then, there exists a set $\Omega \subset I$, with total measure, such that $\forall t \in \Omega$ and $\forall n \in \mathbb{N}$, $\beta_n(t) \geq \lambda$. Given $\varepsilon > 0$, it follows that,
\begin{multline*}
\mu (\{t \in (0,T): -\beta(t) + \lambda > \varepsilon \}) =\mu (\{t \in (0,T): -\beta(t) + \lambda > \varepsilon \} \cap \Omega) \\
\leq \mu (\{t \in (0,T): \mid \beta_n (t)-\beta(t)  \mid \geq \varepsilon \})
\leq \|\beta_n - \beta \|_{H^1(I)}^2/\varepsilon ^2,
\end{multline*}
Where the Tchebychev's inequality and the inclusion $H^1(I)\subset L^2(I)$ were used in the last estimate.
Taking the limit when $n\rightarrow \infty$, $\mu (\{t \in (0,T): -\beta(t) + \lambda > \varepsilon \})=0$, and since $\varepsilon>0$ is arbitrary, we can conclude that $\mu (\{t \in (0,T): \beta(t) - \lambda< 0\})=0$. In other words, $\beta \in D_\lambda(\mathcal{J})$.

Finally, if $\beta_0,\beta_1\in D_{\lambda}(\mathcal{J})$ and $\varepsilon \in [0,1]$, then, $\varepsilon \beta_0 + (1-\varepsilon) \beta_1 \geq \varepsilon \lambda + (1-\varepsilon) \lambda = \lambda$ a.e. and $D_{\lambda}(\mathcal{J})$ is a convex set.
\end{proof}


The following lemma is a useful fact from functional analysis.
\begin{lemma}\label{lem_con_pun}
Let $\{\beta^n\}_{n \in \mathbb{N}}$ be a sequence in $H^1(I)$. Let $\beta \in H^1(I)$ be such that $\beta ^n$ weakly converges to $\beta$ ($\beta_n\rightharpoonup \beta$) in $H^1(I)$. Then, $\lim_{n\rightarrow\infty}\beta ^n(t)=\beta(t)$ for any $t\in I$.
\end{lemma}

The following result is a direct consequence of the previous lemma.
\begin{corollary} \label{Coro_fec_fraco}
$D_{\lambda}(\mathcal{J})$ is weakly closed, $\forall \lambda > 0$.
\end{corollary}


\begin{definition} 
\label{def_f_beta} The function $\boldsymbol F:U \rightarrow \mathbb{R}^9$ with
\[U=(-T_0,T_0)\times B_{\delta_0}(\textit{\textbf{x}}_0) \times D(\mathcal{J})_{\lambda} \subset \mathbb{R} \times \mathbb{R}^9 \times H^1(I),\]
where $T_0 > T$, $\lambda\in[0,1/2)$, and $\delta_0 = 2 \|\textit{\textbf{x}}_0\|_1+1$. Thus,
\[\boldsymbol F(t,\boldsymbol x,\beta)= (F_1(t,\boldsymbol x,\beta), ...,F_9(t,\boldsymbol x,\beta)),\quad (t,\boldsymbol x,\beta))\in U,\] 
with the $F_i$'s given by the RHS of the ODE system in Eqs.~\eqref{model_1}--\eqref{model_2}, with $\beta=\beta_M$ in Eq.~\eqref{eq_betas}.

\end{definition}
The ODE system in Eqs.~\eqref{model_1}--\eqref{model_2} can be written as:
\begin{equation}
\label{re_edo_beta}
\left\{ 
\begin{matrix} 
\boldsymbol x^\prime=\boldsymbol F(t,\boldsymbol x,\beta), \quad t\in I,\\ 
\boldsymbol x(0)=\boldsymbol x_0. \end{matrix} \right.
\end{equation}

\begin{definition} \label{def_J} The operator $\mathcal{J}: D_{\lambda}(\mathcal{J}) \subset X \rightarrow Y$ maps the function $\beta$ to $\mathcal{J}(\beta)= \boldsymbol u$, where $\lambda >0$, $\boldsymbol u$ is the solution of the ODE system in Eq.~\eqref{re_edo_beta}. 
\end{definition}

Taking the assumption that $\beta$ is continuous, Propositions~\ref{pr6} and \ref{pr7} imply the existence and uniqueness of the solution of the ODE problem in Eq.~\eqref{re_edo_beta}. Moreover, the solution is continuously differentiable in $I$. Thus, we have the following lemma:
\begin{lemma}\label{lem2}
The operator $\mathcal{J}: D_{\lambda}(\mathcal{J}) \subset X \rightarrow Y$  is well defined.
\end{lemma}

\begin{lemma}\label{pr11}
For any $\lambda>0$, the operator $\mathcal{J}: D_{\lambda}(\mathcal{J}) \subset X \rightarrow Y$ is injective.
\end{lemma}
\begin{proof}
Let $\beta,\widehat{\beta}\in D_{\lambda}(\mathcal{J})$ be such that $\mathcal{J}(\beta)=\mathcal{J}(\widehat{\beta})$. 
Denoting $\boldsymbol u=\mathcal{J}(\beta)=\mathcal{J}(\widehat{\beta})$, it follows that, 
%
%
$\beta(c u_{4} + u_{5}+ a u_{6} + b u_{7}) = \widehat{\beta}(c u_{4} + u_{5}+ a u_{6} + b u_{7})$.

Since $\beta$ is strictly positive in $I$, by Propositions~\ref{lem3} and \ref{pr4}, and Corollary~\ref{cor1}, 
$c u_{4} + u_{5}+ a u_{6} + b u_{7}$ is strictly positive in $(0,T]$, which implies that $\beta = \widehat{\beta}$ in $(0,t]$. Also, by continuity, $\beta(0)=\widehat{\beta}(0)$. Thus, $\beta = \widehat{\beta}$ in $I$.
\end{proof}

\begin{lemma}\label{lem_cont_unif}
Let $\{\beta^n\}_{n\in \mathbb{N}}$ be a sequence in $D_\lambda(\mathcal{J})$ that converges to $\beta \in (\mathcal{J})$ with respect to the $C(\overline{I})$ norm. Defining $\boldsymbol F_n(t,\boldsymbol x)= \boldsymbol F(t,\boldsymbol x,\beta_n)$ and $\overline{\boldsymbol F}(t,\boldsymbol x)= \boldsymbol F(t,\boldsymbol x,\beta)$, with $(t,\boldsymbol x)\in U_0=[0, T_0] \times \overline{B}_{\delta_0}(\textbf{x}_0)$, then $\lim_{n \to \infty} \|\boldsymbol F^n -\overline{\boldsymbol F}\|_{C(U_0)^9}=0$.

\end{lemma}
\begin{proof}
Let $(t,\boldsymbol x)$ be an arbitrary element of $U_0$. 
Using $\beta^n_M=\beta^n$ and $\beta_M=\beta$, the other transmission rates are defined as in Eq.~\eqref{eq_betas}.
Take $K=\max\{a,b,c,1\}$ and note that:
\begin{multline*}
|F^n_1(t,\boldsymbol x)-\overline{F}_1(t,\boldsymbol x)|
%
%
\leq x_1[|\beta^n_A (t)-\beta_A(t)|x_{4} + |\beta^n_M(t)-\beta_M(t)|x_{5}\\ + |\beta^n_S(t)-\beta_S(t)|x_{6} + |\beta^n_C(t)-\beta_C(t)|x_{7}]
\leq 4K\delta_0^2\|\beta^n -\beta\|_{C(I)}.
\end{multline*}
Analogously, $|F^n_3 (t,\boldsymbol x)-\overline{F}_3(t,\boldsymbol x)| \leq 4 \delta_0^2K \|\beta^n- \beta \|_{C(I)}$. Moreover, for $i=2,4,5,6,7,8,9$, $F_i(t,\boldsymbol x,\beta^n)= F_i(t,\boldsymbol x,\beta)$ for all $n\in\mathbb{N}$. 
Thus, since $(t,\boldsymbol x)\in U_0$ was arbitrary in the above estimates, it follows that
\[\|\boldsymbol F^n-\boldsymbol F\|_{C(U_0)^9} = \max_{i=1,\cdots,9}\|F^n_i- \overline{F}_i\|_{C(U_0)}\leq 4\sqrt{2}K\delta_0^2 \|\beta^n-\beta \|_{C(I)}.\]

Since $\lim_{n \to \infty}\|\beta^n -\beta\|_{C(I)} = 0$, we conclude that $\lim_{n \to \infty} \|\boldsymbol F^n -\boldsymbol F \|_{C(U_0)^9}=0$.
\end{proof}

\begin{lemma}\label{lem_cont_L2}
Let $\{\beta^n\}_{n\in \mathbb{N}}$ be a sequence in $D_\lambda(\mathcal{J})$ that converges in $L^2(I)$ to $\beta \in D(\mathcal{J})_\lambda$. 
Define $\boldsymbol u^n=\mathcal{J}(\beta ^n)$, for all $n\in\mathbb{N}$ and $\boldsymbol u=\mathcal{J}(\beta)$. If $\lim_{n \to \infty}\|u^n_{i} -u_{i}\|_{L^2(I)}=0$, then $\lim_{n \to \infty}\|(u^n_{i})^\prime -u_{i}^\prime\|_{L^2(I)}=0$.
\end{lemma}
\begin{proof}
Let $t \in I$. Denote $\beta^n_A$, $\beta^n_M(=\beta^n)$, $\beta^n_S$ and $\beta^n_C$ as the corresponding rates associated with $\beta^n$, $\forall n \in \mathbb{N}$ and define $K=\max\{a,b,c,1\}$. Thus, applying Jensen's inequality and the fact that $N=\|\boldsymbol u_n\|_1 = \|\boldsymbol{u}\|_1=\|\boldsymbol{x}_0\|_1$, the following estimate holds, 
\begin{multline*}
|u^n _4(t) \beta^n_M(t) - u_4(t)\beta_M(t)|^2
%
%
%
%
\leq (u^n _4(t)|\beta^n_M (t)- \beta_M (t)|+ K\|\beta\|_{C(I)}|u^n_4(t) -u_4 (t)|)^2\\
\leq 2 \|\textit{\textbf{x}}_0\|_1^2 |\beta^n_M (t)- \beta_M (t)|^2 + 2 K^2 \|\beta\|^2_{C(\overline{I})}|u^n_4(t) -u_4 (t)|^2.
\end{multline*}
Integrating from with respect to $t$, from $0$ to $T$, it follows that,
%
%
%
\[\|u^n _4 \beta^n_M - u_4\beta_M \|_{L^2(I)}^2  \leq 2\|\textit{\textbf{x}}_0\|^2_1\|\beta^n- \beta \|_{L^2(I)}^2  + 2K^2\|\beta\|^2_{C(I)}\|u^n_4 -u_4\|_{L^2(I)}^2.\]
Therefore, using the hypotheses, $\lim_{n \to \infty} \|u^n _4 \beta^n_M - u_4\beta_M \|_{L^2(I)}=0$. 
Making similar calculations, it is easy to obtain
$\lim_{n \to \infty} \|u^n_1 u^n _4 \beta^n_M - u_ 1 u_4\beta_M \|_{L^2(I)} =0$ and analogous estimates for the other terms arising in the difference of the derivatives $u_1^\prime$ and $(u^n_1)^\prime$. Thus, 
by the Minkowski inequality, it follows that $\lim_{n \to \infty} \|(u^n_{1})^\prime -u_{1}^\prime\|_{L^2(I)}=0.$
Similarly, it is possible to conclude that, $\lim_{n \to \infty} \|(u^n_ {i})^\prime-u_{i}^\prime\|_{L^2(I)}=0$, for $i =2,\cdots,9$.
\end{proof}

\begin{proposition} \label{pr10}
The operator $\mathcal{J}:D_{\lambda}(\mathcal{J}) \rightarrow H^1(I)^9$ is continuous in the $H^1(I)$-norm topology, for any $\lambda > 0$.
\end{proposition}
\begin{proof}
Let $\{\beta^n\}_{n\in \mathbb{N}}$ be a sequence in $D_{\lambda}(\mathcal{J})$ that converges in the $H^1(I)$-norm to $\beta\in D_{\lambda}(\mathcal{J})$.  Define $\boldsymbol u^n=\mathcal{J}(\beta ^n)$ and $\boldsymbol u=\mathcal{J}(\beta)$. 
By the Sobolev's inequality \cite[p.~212]{brezis2011},  
it follows that $\lim_{n \to \infty}\|\beta^{n} -\beta\|_{L^{\infty}(I)}=0$. Since $\beta^n$ and $\beta$ are continuous, $\{\beta^n\}$ converges uniformly to $\beta$. By Lemma~\ref{lem_cont_unif}, $\{\boldsymbol F(\cdot,\boldsymbol{\cdot},\beta_n)\}_{n\in\mathbb{N}}$ converges uniformly to $\boldsymbol F(\cdot,\boldsymbol{\cdot},\beta)$. 
It is not difficult to show that, for $i=1,\cdots,9$,
\[\|u^n_i-u_i\|_{C(I)}\leq T\|F_i(\cdot,\boldsymbol{\cdot},\beta_n)-F(\cdot,\boldsymbol{\cdot},\beta)\|_{C(I)}.\]
Then, $\{\boldsymbol u^n\}$ converges uniformly to $\boldsymbol u$, and this implies in the $L^2(I)$-convergence. 
 
Since $\lim_{n \to \infty} \Vert \beta^n - \beta \Vert _ {H^1(I)} = 0$, 
by Lemma \ref{lem_cont_L2}, $\lim _{n \to \infty} \|(u^n_{i})^\prime -u_i^\prime\|_{L^2(I)}=0$, with $i=1,\cdots,9$. Then, $\lim _{n \to \infty} \|u^n_{i}-u_i\|_{H^1(I)}^2=0$, with $i=1,...,9$.
%
\end{proof}

\begin{proposition}\label{pro_IMP}
The operator $\mathcal{J}:D_{\lambda}(\mathcal{J}) \rightarrow H^1(I)^9$ is sequentially closed with respect to the weak topologies of X and Y.
\end{proposition}
\begin{proof}
Let $\{\beta^n\}_{n\in\mathbb{N}}$ be a sequence in $D_\lambda(\mathcal{J})$ weakling converging to $\beta$ in $X=H^1(I)$.
Suppose that 
$\{\mathcal{J}(\beta^n)\}_{n\in\mathbb{N}}$ weakly converges to some $\boldsymbol u$ in $H^1(I)^9$. By Corollary~\ref{Coro_fec_fraco}, $\beta \in D_\lambda(\mathcal{J})$. 
Define $\boldsymbol u^n=\mathcal{J}(\beta^n)$. Since $\boldsymbol u^n \rightharpoonup \boldsymbol u$ in $H^1(I)^9$, 
by Lemma \ref{lem_con_pun}, $u_i^n$ converges pointwise to $u_i$ in $I$. By the same lemma, it follows that $\beta_n$ also converges pointwise to $\beta$ in $I$. 
%
%




By the Sobolev's inequality \cite[p.~212]{brezis2011}, 
%
and \cite[Theorem 14.2]{bachman}, 
there exist $C,K> 0$ such that $\|\beta^n\|_{H^1(I)} \leq K$, $\forall n \in \mathbb{N}$ and 
$\|\beta^n\|_{L^{\infty}(I)} \leq C K$, $\forall n \in \mathbb{N}$. Analogously, there exist  $K_i> 0$, with $i=1,\cdots,9$, such that $\|u^n_i\|_{L^{\infty}(I)} \leq C K_i$, $\forall n \in \mathbb{N}$ and $i=1,\cdots,9$. This implies that the sequence $\{F_i(\cdot,\boldsymbol{\cdot},\beta^n)\}_{n\in\mathbb{N}}$ is uniformly bounded in $I$. Thus, for any $t\in (0,T]$, the Dominated Convergence Theorem implies that
%



%
\[\lim_{n \to \infty} \int_0 ^t F_i(s,\boldsymbol u^n(s),\beta_n(s)) ds = \int_0 ^t F_i(s,\boldsymbol u(s),\beta(s)) ds \quad (i=1,\cdots,9).\]

Taking the limit $n\rightarrow \infty$ in the formula $u^n_i(t)= u^n_i(0) + \int_0 ^t F_i(s,\boldsymbol u^n(s),\beta_n(s)) ds$, it follows that $u_i(t)= u_i(0) + \int_0^t F_i(s,\boldsymbol u(s),\beta(s))ds$, with $i=1,\cdots,9$. Then, $u_i^\prime(t)=F_i(t,\boldsymbol u(t),\beta(t))$, $\forall t \in I$ and $i=1,\cdots,9$. 
Moreover, it follows that $\boldsymbol u(0)=\textit{\textbf{x}}_0$, by the pointwise convergence. Therefore, $\boldsymbol u$ is a solution of the ODE system in Eq.~\eqref{def_f_beta} and $u=\mathcal{J}(\beta)$.
\end{proof}

\begin{proposition}\label{pr_DF_X}
The function $\boldsymbol F$ arising in the ODE system in Eq.~\eqref{re_edo_beta} 
is differentiable with respect to $\beta$ and $\boldsymbol x$. The derivatives $\boldsymbol F^\prime_\beta(t,\boldsymbol x, \beta) \in \mathcal{L}(H_1(I),\mathbb{R}^9)$ and $\boldsymbol F'_{\boldsymbol x}(t,\boldsymbol x, \beta) \in \mathcal{L}(\mathbb{R}^9)$ are continuous with respect to $(t,\boldsymbol x, \beta)$.
\end{proposition}

\begin{proof}
Let $\beta$ be in $D_{\lambda}(\mathcal{J})$, $h\in H^1(I)$, and let $\varepsilon >0$ be such that $\beta +\varepsilon h$ be in the interior of $D_{\lambda}(\mathcal{J})$, denoted by $D_{\lambda}(\mathcal{J})^\circ$. 
\begin{multline*}
F_{1}(t,\boldsymbol x,\beta + \varepsilon h)- F_{1}(t,\boldsymbol x,\beta)=
-x_{1} (\beta +\varepsilon h) (c x_{4}+  x_{5}+ a x_{6} + b  x_{7}) - \nu x_1\\
+ x_{1} \beta (c x_{4}+  x_{5}+ a x_{6} + b  x_{7})+\nu x_1
=-x_{1} \varepsilon h (c x_{4}+  x_{5}+ a x_{6} + b  x_{7}).
\end{multline*}
Thus, taking the limit $\varepsilon\rightarrow 0$, it follows that
\[\lim_{\varepsilon\rightarrow 0}\frac{F_{1}(t,\boldsymbol x,\beta + \varepsilon h)- F_{1}(t,\boldsymbol u,\beta)}{\varepsilon} = -x_{1} (c x_{4}+  x_{5}+ a x_{6} + b  x_{7}) h.\]

The corresponding limit for $F_{3}$ is similar. The corresponding limits are zero for the other $\boldsymbol F$ components since they do not depend on $\beta$. Therefore,
\begin{multline*}
    \boldsymbol F'_{\beta}(t,\boldsymbol x,\beta)(h)=\lim_{\varepsilon\rightarrow 0}\frac{\boldsymbol F(t,\boldsymbol x,\beta + \varepsilon h)- \boldsymbol F(t,\boldsymbol x,\beta)}{\varepsilon}\\
=(-x_{1}(c x_{4}+  x_{5}+ a x_{6} + b  x_{7}),0,x_{1}(c x_{4}+  x_{5}+ a x_{6} + b  x_{7}), 0,0,0,0,0,0)h.
\end{multline*}

Thus, $\boldsymbol F$ is Gateaux differentiable at every $\beta$ in $D_{\lambda}(\mathcal{J})^{\circ}$. It is not difficult to show that $\boldsymbol F'_{\beta}(t,x, \beta)$ is sequentially continuous with respect to $(t,x,\beta)$ and it is a continuous linear operator in $H^1(I)$. 

The assertion about the derivative $\boldsymbol F^\prime_{\boldsymbol x}$ is an immediate consequence of the definition of $\boldsymbol F$ and differentiability results from Vector Calculus \cite{cartan1971}.
\end{proof} 

Heuristically, for $\beta \in D_{\lambda}(\mathcal{J})$, $\boldsymbol u=\mathcal{J}(\beta)$, and $h \in X$,
 the ODE system associated with the directional derivative of the operator $\mathcal{J}$, in the direction $h$, can be written as:
 \begin{multline}
    \frac{dy_{1}}{dt} = -y_{1}(\beta_A u_{4} + \beta_M u_{5}+ \beta_S u_{6} + \beta_C u_{7})\\
    -u_{1}(\beta_A y_{4} + \beta_M y_{5}+ \beta_S y_{6} + \beta_C y_{7})
- h u_1(cu_4 + u_5 + au_6 +b u_7) -\nu y_1\label{eq_DJ_1}
\end{multline}
\begin{align}
    \frac{dy_{2}}{dt} = \nu y_{1}
\end{align}
\begin{multline}
    \frac{dy_{3}}{dt} =y_{1}(\beta_A u_{4} + \beta_M u_{5}+ \beta_S u_{6} + \beta_C u_{7})+u_{1}(\beta_A y_{4} + \beta_M y_{5}+ \beta_S y_{6} + \beta_C y_{7})\\ + h u_1(cu_4 + u_5 + au_6 +b u_7)  - \sigma y_{3}
\end{multline}
\begin{gather}
    \frac{dy_{4}}{dt} = (1-p)\sigma y_{3} - \gamma_{R,A} y_{4}\\
    \frac{dy_{5}}{dt} = p\sigma y_{3} - (\gamma_{R,M} + \alpha_S) y_{5}\\
    \frac{dy_{6}}{dt} = \alpha_S y_{5} - (\gamma_{R,S} + \alpha_C) y_{6}\\
    \frac{dy_{7}}{dt} = \alpha_C y_{6} - (\gamma_{R,C} + \delta_D) y_{7}\\
    \frac{dy_{8}}{dt} =\gamma_{R,A} y_{4} + \gamma_{R,M} y_{5} + \gamma_{R,S} y_{6} + \gamma_{R,C} y_{7}\\
    \frac{dy_{9}}{dt} = \delta_D y_{7}\label{eq_DJ_2}
\end{gather}
Thus, using the notation of Prop.~\ref{pr_DF_X}, the ODE system in Eqs.~\eqref{eq_DJ_1}--\eqref{eq_DJ_2} rewritten in a compact form as follows.
\begin{equation}
\label{eq_DJ}
\left \{ \begin{matrix} \boldsymbol y'=\boldsymbol F'_{x}(t,\boldsymbol u(t),\beta)y+\boldsymbol F'_{\beta}(t,\boldsymbol u(t),\beta)h, \quad t\in I, \\ 
\boldsymbol y(0)=\boldsymbol 0. \end{matrix} 
\right.
\end{equation}
 
\begin{definition} \label{def_DJ} Let $\beta \in D_{\lambda}(\mathcal{J})$. The operator $\mathcal W(\beta): X \rightarrow Y$ maps $h\in X$  onto $\mathcal W(\beta)(h)= \boldsymbol v$, the solution of the ODE system in Eq.~\eqref{eq_DJ}.
\end{definition}
%
%

The arguments used in the proof of the following theorem are similar to those in \cite[Theorem 3.6.1]{cartan1971}.

\begin{theorem}\label{T_DJ}
The operator $\mathcal{J}$ is Fréchet differentiable in the interior of $D_\lambda(\mathcal{J})$ and its Fréchet derivative is the operator $\mathcal W$ of Definition~\ref{def_DJ}.
\end{theorem}

\begin{proof}
Let $\beta \in D_{\lambda}(\mathcal{J})^\circ$ and $h \in X$ be such that $\beta+h \in D_{\lambda}(\mathcal{J})$. Set $\boldsymbol u^h = \mathcal{J}(\beta+h)$ and  $\boldsymbol u=\mathcal{J}(\beta)$. 
Consider the following ODE system, which is defined in $U=B_{\delta_0}(\boldsymbol{x}_0)\times D_\lambda(\mathcal{J})^{\circ}$,
\begin{equation}
 \left \{ \begin{matrix} \boldsymbol x^\prime=\boldsymbol F(t,\boldsymbol x,\beta), \quad t\in I, \\ 
 y^\prime=0, \quad t \in I,\\
(\boldsymbol x(0),y(0))=(\boldsymbol{x}_0, \beta). \end{matrix} \right.
\label{eq:sysA}
\end{equation}
Defining $\boldsymbol G(t,(\boldsymbol x,y))=(\boldsymbol F(t,\boldsymbol x,y),0)$, the ODE system in Eq.~\eqref{eq:sysA} is equivalent to
\begin{equation}
 \left \{ \begin{matrix} \boldsymbol (\boldsymbol x^\prime,y^\prime)=\boldsymbol G(t,(\boldsymbol x,y)), \quad t\in I, \\ 
(\boldsymbol x(0),y(0))=(\boldsymbol{x}_0, \beta). \end{matrix} \right.
\end{equation}
Now, consider the variation equation below,
\begin{equation}\label{eq_DW}
 \left \{ \begin{matrix} \boldsymbol\omega^\prime= A\boldsymbol\omega, \quad t\in I, \\ 
\boldsymbol\omega(0)=Id_{E}, \end{matrix} \right.
\end{equation}
with
\[
A=
\left[ {\begin{array}{cccc}
  \boldsymbol F^\prime_{x}(t,\boldsymbol u(t),\beta)& \boldsymbol F^\prime_{\beta}(t,\boldsymbol u(t),\beta) \\
  0 & 0 \\
\end{array} } \right],
\]
and $Id_E$ the identity operator in $E = \mathbb{R}^9 \times H^1(I)$. Then, $A(t)=\boldsymbol G^\prime_{(\boldsymbol x,y)}(t,\boldsymbol u(t),\beta)$, $\forall t \in I$. 

Combining the continuity of 
$\boldsymbol F^\prime_{\boldsymbol x}(t,\boldsymbol x,\beta)$ w.r.t. $(t,\boldsymbol x,\beta)$, by Proposition~\ref{pr_DF_X}, and the continuity of $\boldsymbol u(t)$, 
$\boldsymbol F^\prime_{\boldsymbol x}(t,\boldsymbol u(t),\beta)$ 
 is continuous in $t$. Similarly, $t\mapsto\boldsymbol F^\prime_{\beta}(t,\boldsymbol u(t),\beta)$ 
is continuous. 
Then, since $I$ is compact, $\displaystyle K=\sup_{t \in I} \|A(t)\|$ is finite and, consequently, $A(t)\boldsymbol \omega $ is K-Lipchitz in $\boldsymbol \omega$. Thus, there exists a unique solution of the ODE problem in Eq.~\eqref{eq_DW} for $t\in I$ and defined in $\mathcal{L}(E)$. 
In particular, 
$\boldsymbol \omega'(t)(\boldsymbol 0,h)= A(t)\omega(t)(\boldsymbol 0,h)$,
which means that, for all $t\in I$, it follows,
\begin{gather*}
\boldsymbol \omega'_1(t)(\boldsymbol 0, h)=\boldsymbol F'_{\boldsymbol x}(t,\boldsymbol u(t),\beta) \boldsymbol \omega_1(t)(\boldsymbol 0, h) + \boldsymbol F'_{\beta}(t,\boldsymbol u(t),\beta)\omega_2(t)(\boldsymbol 0, h),\\
\omega'_2(t)(\boldsymbol 0, h)=0.
\end{gather*}
Thus, $\omega_2(t)=Id_{H^1(I)}$, $\forall t \in I$ since $\omega'_2 (t)=0$ and $\omega_2(0)=Id_{H^1(I)}$. 
Moreover, by defining $\boldsymbol v(t) = \boldsymbol \omega_1(t)(\boldsymbol 0,h)$, it follows that,
\begin{gather*}
\boldsymbol v'(t)=\boldsymbol F'_{\boldsymbol x}(t,\boldsymbol u(t),\beta)\boldsymbol v(t) + F'_{\beta}(t,\boldsymbol u(t),\beta)h,\\
%
%
\boldsymbol v(0)=\boldsymbol \omega_1(0)(\boldsymbol 0,h)=[Id_{\mathbb{R}^9},0](\boldsymbol 0,h)=\boldsymbol 0,
\end{gather*}
and thus,$\boldsymbol v$ is the solution of the ODE problem in Eq.~\eqref{eq_DJ}.

If $\boldsymbol z(t)=(\boldsymbol u^h(t),\beta + h)- (\boldsymbol u(t),\beta)- \boldsymbol \omega(t)(0,h)$, then, 
\begin{gather*}
 \boldsymbol z'(t) = ((\boldsymbol u^{h})^{\prime} (t)- \boldsymbol u^\prime(t),0)- A(t) \boldsymbol \omega(t)(0,h),\\
 A(t)\boldsymbol z(t) = A(t)(\boldsymbol u^h(t)-\boldsymbol u(t),h)- A(t) \boldsymbol \omega(t)(0,h).
\end{gather*}
%
Now, combining the definitions of $\boldsymbol F$, $\boldsymbol u^h$, and $\boldsymbol G$, it follows that,
\begin{align*}
\boldsymbol z'(t) - A(t)\boldsymbol z(t)
& = (\boldsymbol F(t,\boldsymbol u^h(t),\beta +h),0)- (\boldsymbol F(t,\boldsymbol u (t),\beta),0) - A(t)(\boldsymbol u^h(t)-\boldsymbol u(t),h)\\
& = \boldsymbol G( t,\boldsymbol u^h(t),\beta +h) - \boldsymbol G(t,\boldsymbol u (t),\beta) - \boldsymbol G_{(x,y)} '(t,\boldsymbol u (t),\beta)(\boldsymbol u^h(t)-\boldsymbol u(t),h)
\end{align*}
Define for $\lambda \in [0,1]$ the function 
$$\phi (\lambda) = \boldsymbol G\left(t,\lambda(\boldsymbol u^{h} (t),\beta+h)+(1-\lambda)(\boldsymbol u (t),\beta)\right) - \lambda \boldsymbol G_{(x,y)}^\prime\left(t,\boldsymbol u (t),\beta\right)(u^h(t)-u(t),h).$$ 
Then, by the mean value inequality applied to $\phi$ in $[0,1]$, it follows that,
\[\|z^\prime(t) - A(t)z(t)\|_{E} = \|\phi(1)-\phi(0)\|_E
\leq M \|(\boldsymbol u^h(t)-\boldsymbol u(t),h)\|_{E},\]
with $\displaystyle M = \sup_{0\leq \lambda \leq 1} \|\boldsymbol G^\prime_{(x,y)}\left(t,\lambda(\boldsymbol u^{h} (t),\beta+h+(1-\lambda)(\boldsymbol u(t),\beta))\right) -  \boldsymbol G_{(x,y)}^\prime(t,\boldsymbol u (t),\beta)\|_{\mathcal{L}(E)}$.
%

Notice that, it is not difficult to prove that the function
$$\boldsymbol G'_{(x,y)}(t,\lambda (\mathcal{J}(\beta+h) (t),\beta+h)+(1-\lambda)(\mathcal{J}(\beta),\beta))$$
converges uniformly in the interval $I$ to $\boldsymbol G'_{(x,y)}(t,\mathcal{J}(\beta) (t),\beta)$, when $h \to 0$. 
%
Thus, 
given an $\varepsilon >0$, there exists $\eta> 0$, such that, if $\|h\|_{X}<\eta$, then, $M<\varepsilon$ and 
$$\|\boldsymbol z'(t) - A(t)\boldsymbol z(t) \|_{E} \leq \varepsilon \|(\boldsymbol u^h(t)- \boldsymbol u(t),h)\|_{E}, \quad \forall t \in I.$$

Since $A(t)\boldsymbol \omega$ is K-Lipchtiz in $\boldsymbol \omega$ and $(\boldsymbol u^h(t),\beta + h)$ and $(\boldsymbol u(t),\beta)$ are solutions of the ODE $\boldsymbol \omega^\prime= A(t)\boldsymbol \omega$, then, by \cite[Proposition 1.10.1]{cartan1971}, there exists $C>0$, not depending on $h$, such that,
$$\|(\boldsymbol u^h(t)- \boldsymbol u(t),h)\|_{E} < C \|(0,h) \|_{E} = C \|h\|_{X}, \quad \forall t \in I,$$
%
%
and thus,
$$\|\boldsymbol z^\prime(t) - A(t)\boldsymbol z(t) \|_{E} < C \varepsilon \|h\|_{X} , \quad \forall t \in I.$$
In other words, $\boldsymbol z(t)$ is an $\epsilon$--approximation of the ODE $\boldsymbol \omega^\prime = A(t)\boldsymbol \omega$. 
Moreover, it is not difficult to show that $\boldsymbol z(0) = \boldsymbol 0$ and $\widehat{\boldsymbol z}(t)\equiv 0$ is also a solution of $\boldsymbol \omega^\prime = A(t)\boldsymbol \omega$. Then, 
%
by the fundamental lemma \cite{cartan1971},
\[\|z(t)\|_E=\|\boldsymbol z(t)-\widehat{\boldsymbol z}(t)\|_{E} \leq \mbox{e}^{Kt}\|\boldsymbol z(0)-\widehat{\boldsymbol z}(0)\|_{E} + C \varepsilon\dfrac{e^{Kt}-1}{K}\|h\|_{X}\leq \widehat{C} \varepsilon \|h\|_{X},\]
with $\widehat{C}=C(\mbox{e}^{KT}-1)/K$. 
Now, notice that,
\begin{multline*}
\|\boldsymbol z(t)\|_E = \|(\boldsymbol u^h(t),\beta + h) - (\boldsymbol u(t),\beta) - \boldsymbol \omega(t)(\boldsymbol 0,h) \|_E\\
= \|\boldsymbol u^h(t) - \boldsymbol u(t) -\boldsymbol v(t) \|_2\leq \widehat{C}\varepsilon\|h\|_X.
\end{multline*}
Thus, taking the square of the components of the estimate above and integrating it from $0$ to $T$, it follows that
%

%
%
%
\[\|u^h_1 - u_1 -v_1 \|^2_{L^2(I)} +...+ \|u^h_9 - u_9 -v_9 \|^2_{L^2(I)} \leq T \widehat{C}^2\varepsilon^2 \|h\|^2_X.\]

Notice that, $\|z^\prime(t)\|_E \leq \|z^\prime(t)-A(t)z(t)\|_E + K\|z(t)\|_E$. Then, $\|z^\prime(t)\|_E \leq (C+K\widehat{C})\varepsilon\|h\|_X$, and since $\|z^\prime(t)\|_E=\|\boldsymbol (u^{h}_1)^\prime(t) - \boldsymbol u^\prime(t) -\boldsymbol v^\prime(t) \Vert_2$, 
we conclude that 
\[\Vert u^{h \text{ } \prime}_1 - u'_1 -v'_1 \Vert^2_{L^2(I)} +...+ \Vert u^{h \text{ } \prime}_9 - u'_9 -v'_9 \Vert^2_{L^2(I)} \leq T (C+K\widehat{C})^2\varepsilon^2 \Vert h \Vert^2_X.\]
Therefore, 
for any $\varepsilon>0$, there exists an $\eta>0$, such that, if $\Vert h \Vert_X \leq \eta$, then,
\begin{equation}\label{eq_Der_Fre}
\Vert \boldsymbol u^{h} - \boldsymbol u -\boldsymbol v \Vert_Y \leq \sqrt{2T} \widetilde{C}\varepsilon \Vert h \Vert_X,
\end{equation}
where $\widetilde{C} = \max{\widehat{C},C+K\widehat{C}}$. 
By the uniqueness of solutions, it follows that $\boldsymbol{v} = \mathcal{W}(\beta)(h)$ and the assertion follows.

\end{proof}

\section{The Regularized Inverse Problem}\label{sec:inverse}
To provide accurate predictions, the proposed SEIR-like model must be calibrated from the observed data in a stable way. Thus, a natural approach is to apply Tikhonov-type regularization to the so-called inverse problem of estimating $\beta$. As mentioned above, the other parameters are assumed to be known. See Table~\ref{Tab_Par}. 

To be more precise, 
Let $\boldsymbol u$ be an element of the operator range $\mathcal{J}$, denoted as $\mathcal{R}(\mathcal{J})$. Thus, we must find $\beta^\dagger \in D_\lambda(\mathcal{J})$ solving the following equation:
\begin{equation} \label{eq_pr_inv}
\mathcal{J}(\beta^\dagger)=\boldsymbol u.
\end{equation}
Notice that, by Lemma~\ref{pr11}, this problem has a unique solution.

In general, it is impossible to access the exact values of $\boldsymbol u$. The infection notification process is noisy. Generally, only the numbers of infections, hospitalizations, and deaths are available. Thus, we assume that $\boldsymbol u^\delta$ represents the observable quantities, which can be, for example, the daily number of infections, that is also noisy and satisfies:
\begin{equation}\label{eq_dta_ruido}
\Vert \boldsymbol u-\boldsymbol u^\delta \Vert_{(H^1(0,T))^9} \leq \delta.
\end{equation}

To find a stable approximation of the solution of the inverse problem in Eq.~\eqref{eq_pr_inv}, define the corresponding Tikhonov-type functional as follows:
\begin{equation}\label{eq_Tik}
\mathcal{T}_{\alpha,\boldsymbol u^{\delta}}(\beta):= \Vert \mathcal{J}(\beta) - \boldsymbol u^{\delta}\Vert_{Y}^2 + \alpha \Vert \beta - \beta_{0} \Vert_{X}^2
\end{equation}

The following theorems are direct consequences of the results of \cite[Section 3.1]{scherzer2009} and Propositions~\ref{pr_in_Novacio} and \ref{pro_IMP} from Section~\ref{sec_pr_dir}. They guarantee the existence of minimizes for the functional in Eq.~\eqref{eq_Tik}. In addition, by properly choosing the regularization parameter $\alpha$, it is possible to obtain a sequence that approximates the solution of Eq.~\eqref{eq_pr_inv}, stably.

\begin{theorem} (Existence)\label{T_existencia}
Let $\alpha >0$. For any $\boldsymbol u^{\delta}$ in $H^{1}(0,T)^9$, there exists at least one minimizer of $\mathcal{T}_{\alpha, u^{\delta}}$, defined in Eq.~(\ref{eq_Tik}), in $H^{1}(0,T)$.
\end{theorem}

A regularized solution $\beta_k^\delta$ is stable if it continuously depends on $\boldsymbol u^\delta$.

\begin{theorem}(Stability)\label{T_estabilidad}
Let $\alpha > 0$. Given a sequence $\{\boldsymbol u_k\}_{k \in \mathbb{N}}$ such that $\displaystyle\lim_{k \to \infty}\boldsymbol u_k= \boldsymbol u ^{\delta}$. Then the sequence $\{\beta_k\}_{k \in \mathbb{N}}$ has a convergent subsequence, where $\beta_k \in argmin \mathcal{T}_{\alpha, u_k}$, $\forall k \in \mathcal{N}$. Moreover, every convergent subsequence converges to the minimizer of $\mathcal{T}_{\alpha,u^{\delta}}$.
\end{theorem}

\begin{theorem}(Convergence)\label{T_convergencia}
Assume that $\boldsymbol u \in \mathcal{R}(\mathcal{J})$.  Let $\alpha : (0, \infty) \rightarrow (0,\infty)$ be such that:
\[
\lim_{\delta \to 0}\alpha(\delta) = 0 \quad \mbox{and} \quad \lim_{\delta \to 0} \frac{\delta^2}{\alpha(\delta)}=0.
\]
Let $\{\delta_k\}_{k \in \mathbb{N}}$ be a sequence of positive numbers such that $\displaystyle \lim_{k \to \infty}\delta_k = 0$. Assume also that $\boldsymbol u^{\delta_k}$ satisfies Equation~(\ref{eq_dta_ruido}). Then, the sequence $\{\beta_k\}_{k \in \mathbb{N}}$, with $\beta_k  \in argmin \mathcal{T}_{\alpha(\delta_k), \boldsymbol u_k}$ for all $k\in\mathbb{N}$, has a convergent subsequence, 
that converges to $\beta^{\dagger}$.
\end{theorem}
Notice that, as mentioned above, the solution of the inverse problem in Eq.~\eqref{eq_pr_inv} has a unique solution.

\section{Numerical Examples}\label{sec:numerics}
This section aims to estimate the transmission parameter from the observed new infections. All other parameters are assumed to be known. With synthetic data, scenarios with no vaccination and with vaccination are considered. With real data, the daily number of cases in Chicago is considered. 

\subsection{Synthetic Data}
In this set of examples, a population of 270 thousand is considered. At time $t=0$, everyone is susceptible but seven mildly infected individuals. The experiment lasts 300 days. During this period, two massive outbreaks occur. Both have peaks on days $t=27$ and $t=177$, respectively. 

Only the new symptomatic infections are observed. The asymptomatic are not registered. The parameters' values used to generate the daily number of observed infections are in Tab.~\ref{Tab_Par}. The same parameter values are used to calibrate the model. 

The data $u^\delta$ is corrupted by a Gaussian noise as follows
$$
u^\delta(t) = \tilde u(t)\times(1+\delta\epsilon(t)),
$$
where $\tilde u$ is the noiseless data, $\epsilon(t)$ is a standard Gaussian random variable, with independent samples for each $t$, and $\delta$ assumes the values $0$, $0.01$, $0.05$, $0.10$, and $0.20$. 

\begin{figure}[!htb]
     \centering
     \includegraphics[width=0.45\textwidth]{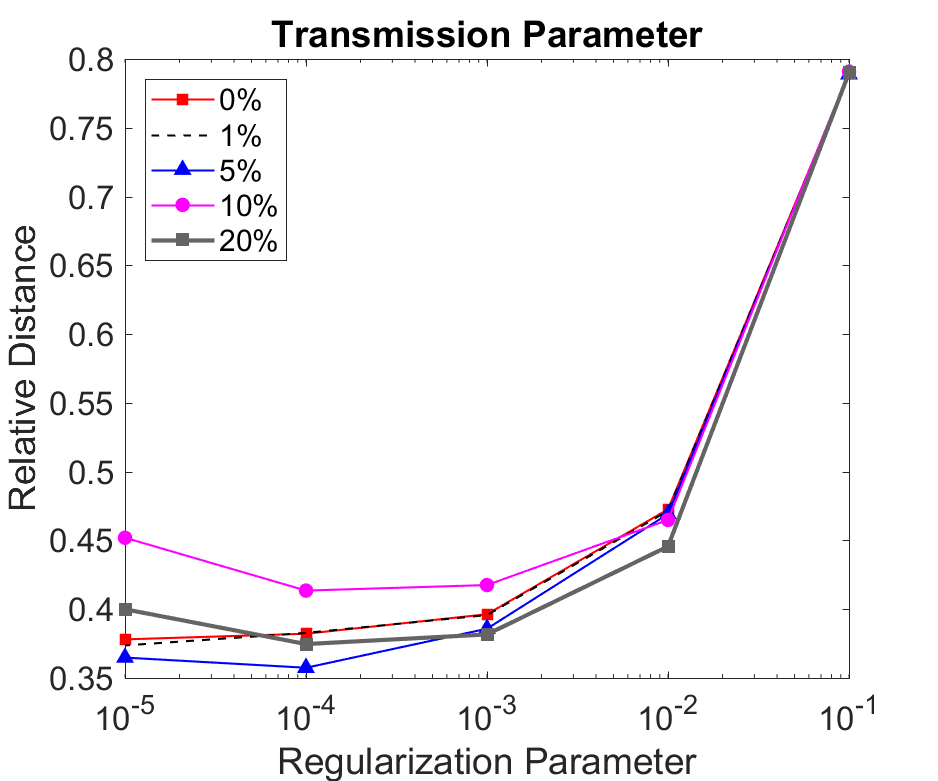}\hfill
     \includegraphics[width=0.45\textwidth]{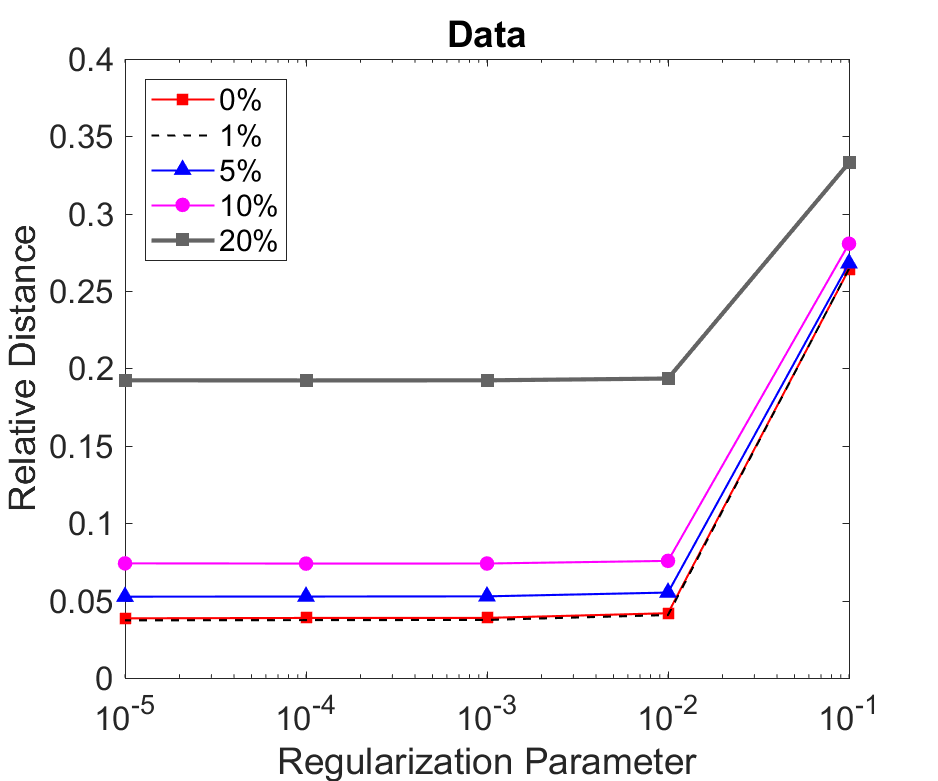}\hfill
     \includegraphics[width=0.45\textwidth]{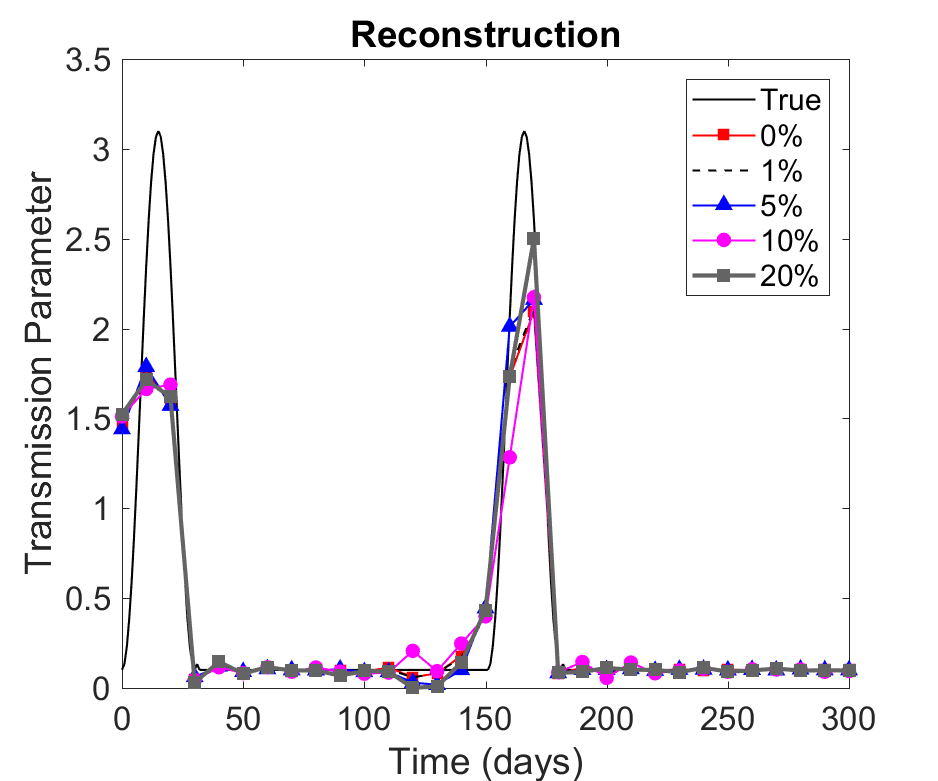}\hfill
     \includegraphics[width=0.45\textwidth]{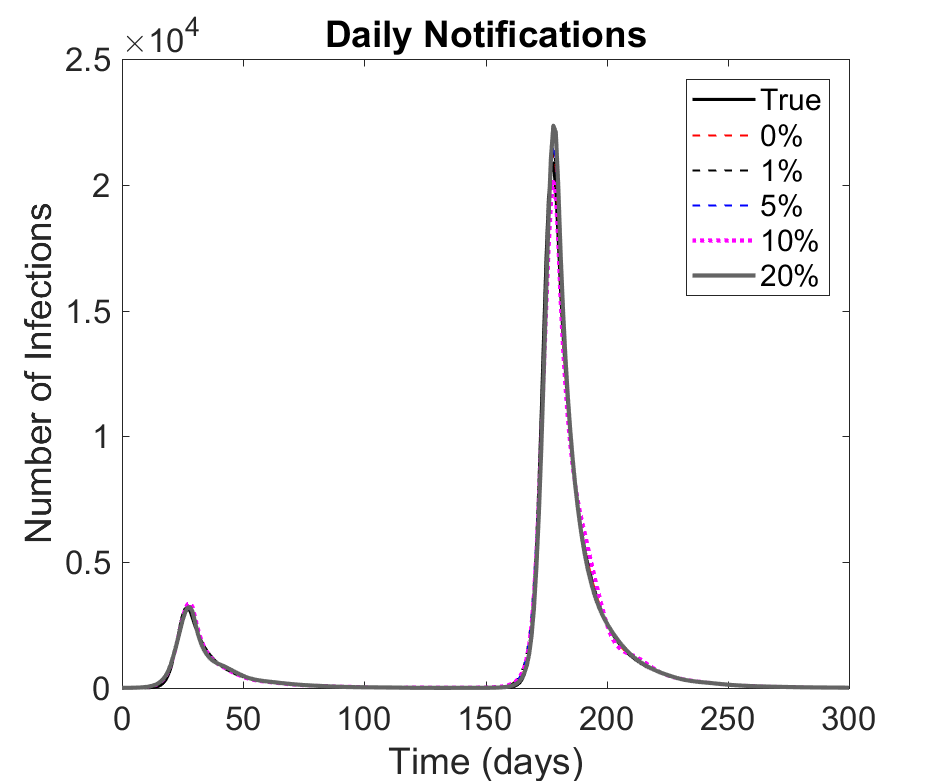}\hfill
     \caption{Top left: Reconstruction normalized error evolution considering different regularization parameter values and noise levels. Top right:  Corresponding evolution of the normalized distance of model predictions to the observed data. Bottom left: Comparison between the reconstructed transmission parameters and the true one (solid black line) for the regularization parameter $\alpha = 10^{-3}$. Bottom right: Comparison of the corresponding model predicted infections with the true number of cases (no noise). No vaccination.}
     \label{fig:results1}
 \end{figure}

In the scenario without vaccination, the vaccination rate is set as $\nu=0$, while, in the scenario with vaccination, it is set as $\nu=0.5/50$ from $t=50$ to $t=100$, otherwise, it is set as $\nu=0$. The transmission parameter used to generate the data is the solid line, in Fig.~\ref{fig:results1}, bottom left panel.

 \begin{figure}[!htb]
     \centering
     \includegraphics[width=0.45\textwidth]{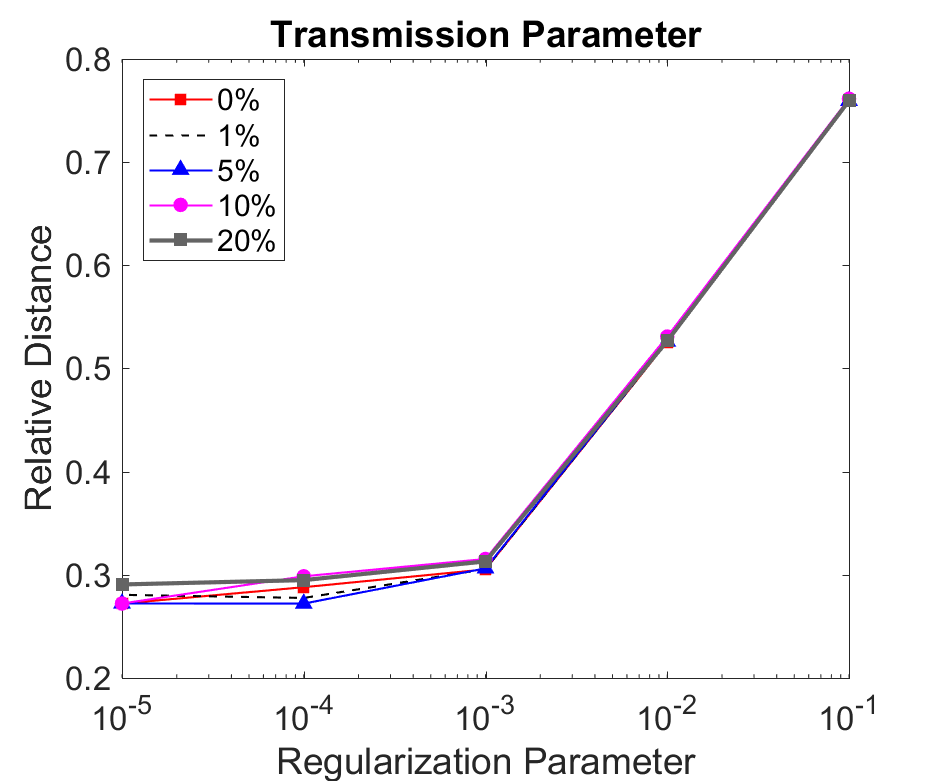}\hfill
     \includegraphics[width=0.45\textwidth]{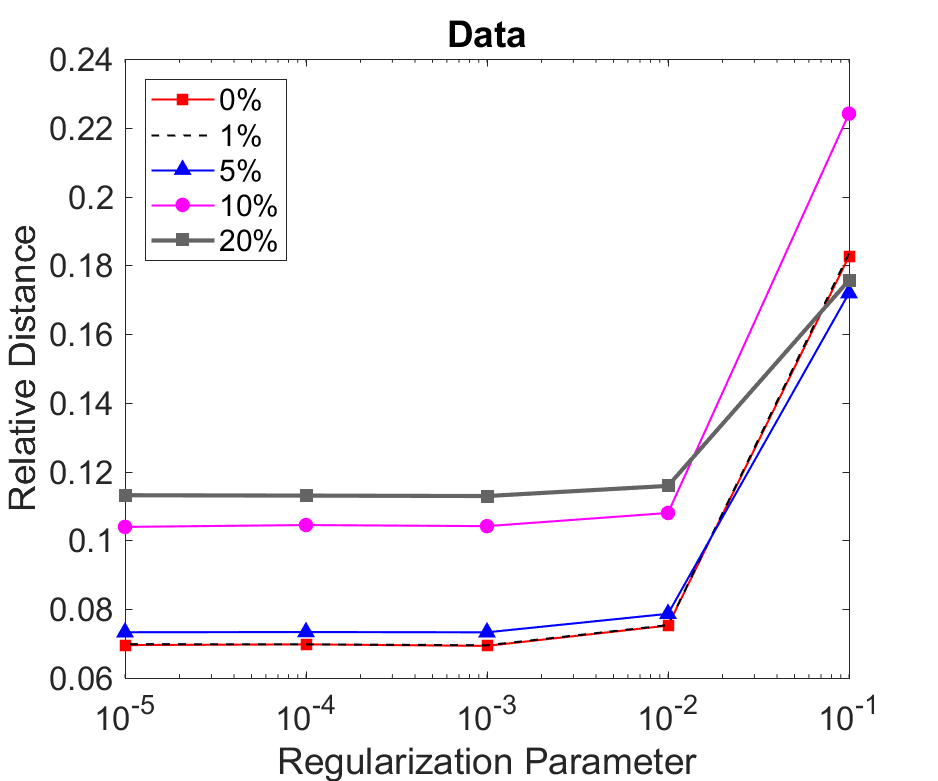}\hfill
     \includegraphics[width=0.45\textwidth]{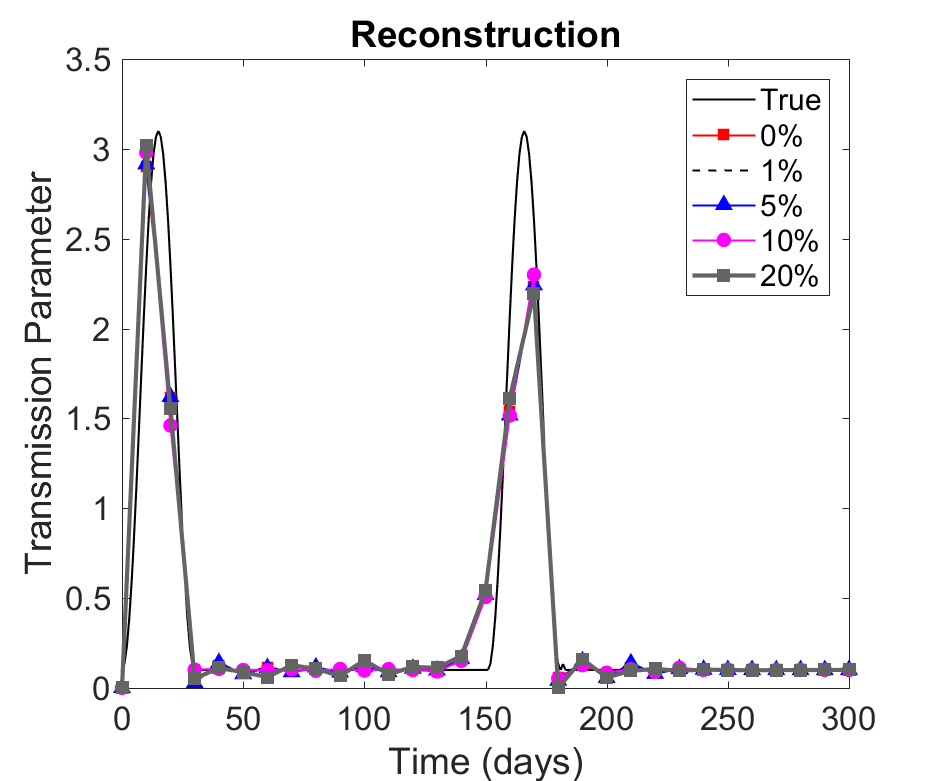}\hfill
     \includegraphics[width=0.45\textwidth]{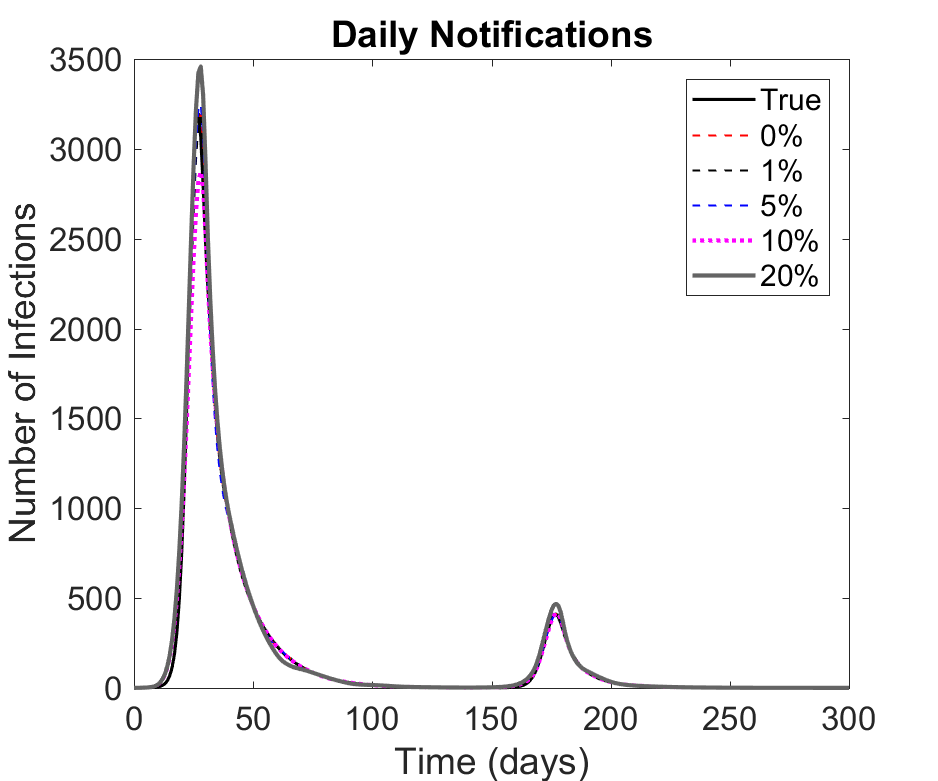}\hfill
     \caption{Top Left: Reconstruction normalized error evolution considering different regularization parameter values and noise levels. Top right: Corresponding evolution of the normalized distance of model predictions to the observed data. Bottom Left: Comparison between the reconstructed transmission parameters and the true one (solid black line) for the regularization parameter $\alpha = 10^{-3}$. Bottom right: Comparison of the corresponding model predicted infections with the true number of cases (no noise). Vaccination.}
     \label{fig:results2}
 \end{figure}

 \begin{figure}[!htb]
     \centering
     \includegraphics[width=0.45\textwidth]{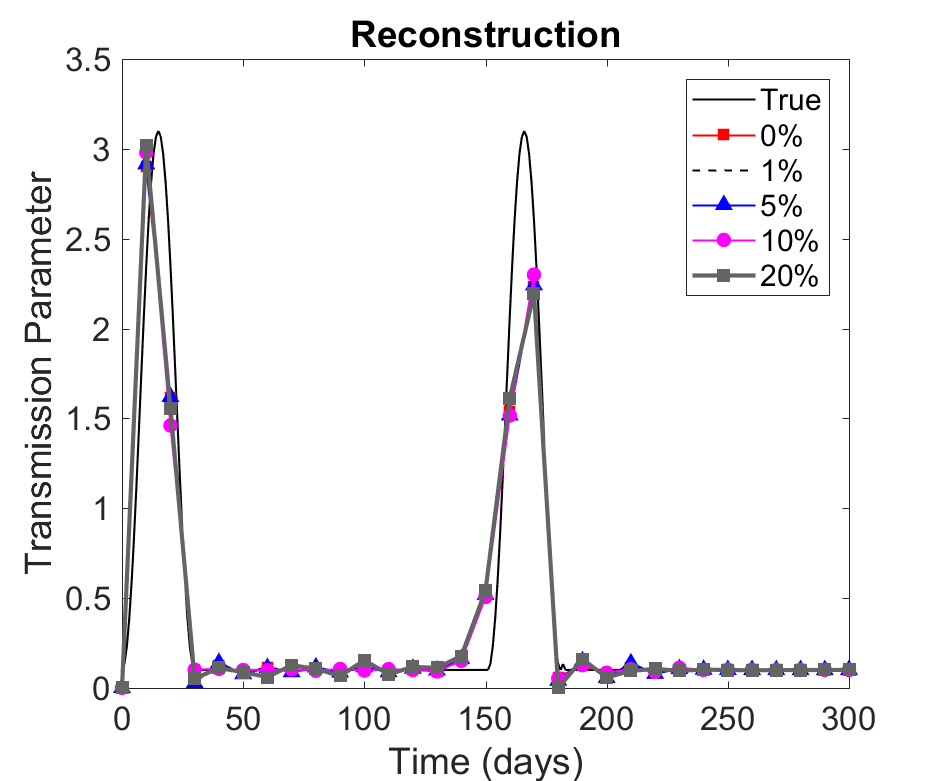}\hfill
     \includegraphics[width=0.45\textwidth]{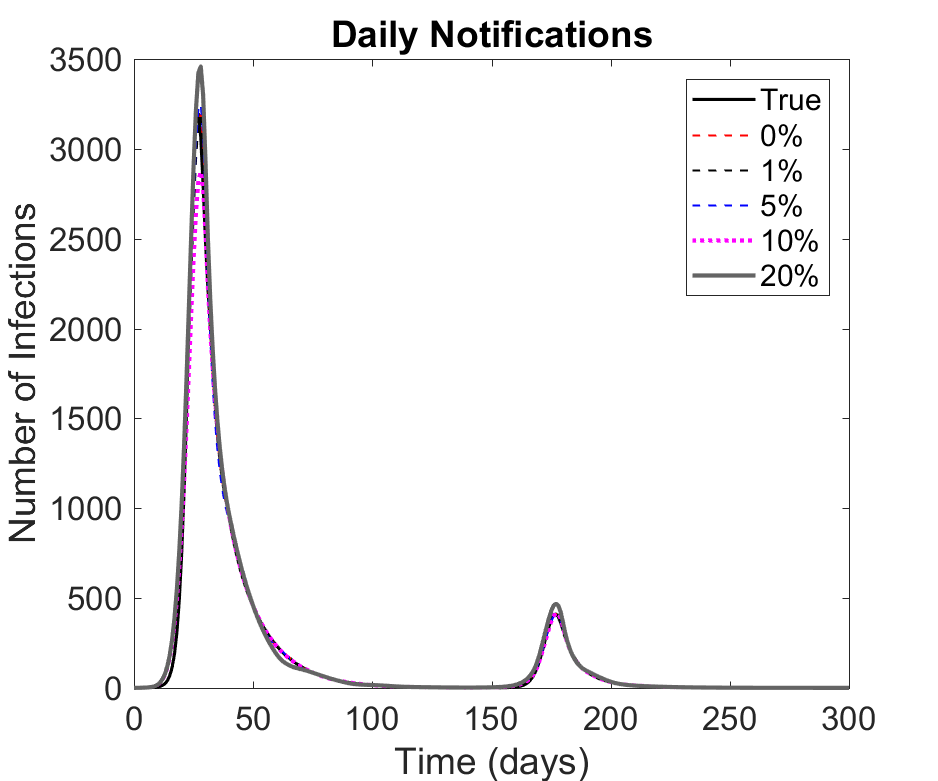}\hfill
     \caption{Left: Comparison between the reconstructed and transmission parameters and the true transmission (solid black line) for the regularization parameter $\alpha = 10^{-3}$. Right: Comparison of the corresponding model predicted infections with the true number of cases (no noise). Wrong vaccination parameter.}
     \label{fig:results3}
 \end{figure}

 To estimate the model, the $H^1(I)$-norm in the data misfit term of the Tikhonov-type functional in Eq.~\ref{eq_Tik} is replaced by the discrete version of the $L^2(I)$-norm. The goodness of fit of the calibrated model is evaluated using the normalized error or normalized distance formula: \[
 \|\hat{Q}-Q^{\mbox{obs}}\|_{2}/\|Q^{\mbox{obs}}\|_{2},
 \]
where $\hat{Q}$ denotes the vector of model predictions or reconstructed values, and $Q^{\mbox{obs}}$ represents the vector of observed quantities or the true parameters. 
To save computational time and avoid an inverse crime, the mesh used to solve the inverse problem is 10 times coarser than the mesh used to generate the data. Since the model uses a step size of one day, the transmission parameter is interpolated linearly.

Figures~\ref{fig:results1}--\ref{fig:results2} present the results without and with vaccination, respectively. It shows the evolutions of the normalized error of reconstructions (top left) and model predictions of daily observed cases (top right), that is, new mild infections. It also compares the reconstructed transmission parameters with the true one (bottom left) and the model predictions of daily observed cases with the noiseless one (bottom right). In the bottom line, only the results that consider the value $\alpha = 10^{-3}$ for the regularization parameter are presented.

To save computational time, the LSQNONLIN default settings were used. For all noise levels, the reconstruction error was similar. In the penalty term, the constant $\beta_0=2$ was used as the prior. It was also used to initialize the minimization algorithm. 

In the no-vaccination scenario, considering all noise levels, the model adhered well to the data. However, for smaller values of the regularization parameter, the prediction error stopped decreasing since the relative distance between two consecutive iterates reached the minimum default value. The precision of the reconstructions was also limited by the mesh used in the estimation since it is ten times coarser than the original one.
In all cases, the model successfully identified the duration and the level of the two peaks in the transmission parameter. It also accurately determined the baseline level, that is, the constant level between outbreaks. In the vaccination scenario, the results were similar, also with adherence to the data and an accurate estimation of the functional transmission parameter.

Figure~\ref{fig:results3} shows the reconstructed transmission parameter considering the wrong vaccination parameters. In the estimation, the parameter $\nu$ is set as 75\% and 125\% of the value used to generate the data. To account for underestimation and overestimation of the vaccine efficacy, respectively. The noise level of the data is 5\% and the regularization parameter is set as $10^{-3}$. In both cases, the estimated transmission was close to the true one, and the model-predicted infections were also adherent to the noiseless data.

These tests illustrate that with synthetic data, the proposed estimation technique is able to correctly reconstruct the original transmission parameter with limited and noisy data, with wrong parameters.

The numerical experiments were implemented in MATLAB. To minimize the Tikhonov-type functional, the LSQNONLIN from MATLAB's optimization toolbox was used. The codes and data sets are available upon request.

\subsection{Real Data}

\paragraph*{Chicago in 2020}
The transmission parameter is estimated from the incidence in Chicago during the first and second waves of COVID-19 in 2020. The data is the daily notifications of COVID-19 infections, from 03-Jan-2020 to 20-Nov-2020. In this period, there was no vaccination. The daily infections time series was smoothed by a 7-day moving average. The mesh used to estimate the transmission parameter has a step size of seven days. To solve the model with a step size of one day, the transmission parameter was linearly interpolated.

 \begin{figure}[!htb]
     \centering
     \includegraphics[width=0.45\textwidth]{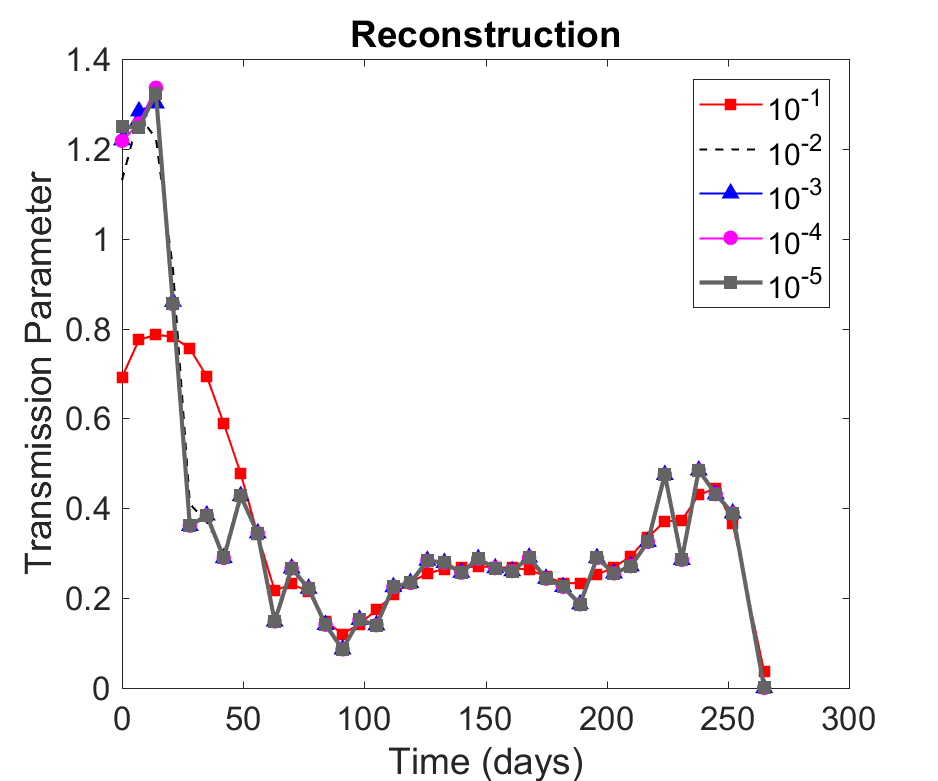}\hfill
     \includegraphics[width=0.45\textwidth]{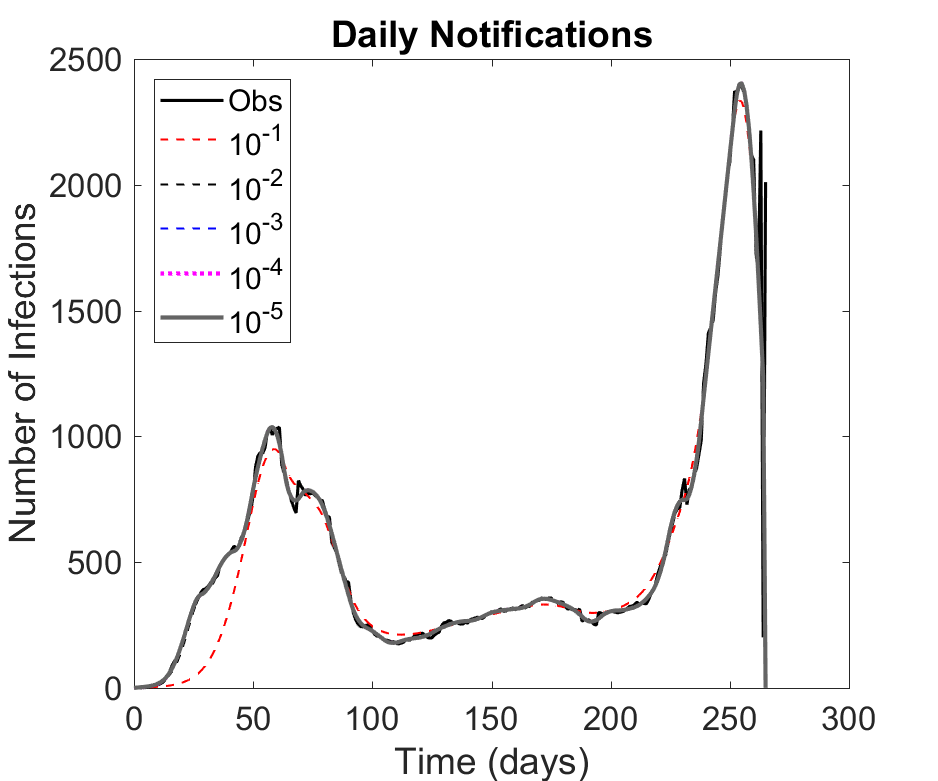}\hfill
     \caption{Left: Reconstructed transmission parameters for different regularization parameter values. Right: Comparison of the corresponding model predictions of cases with the observed daily infections. COVID-19 infections in Chicago, during 2020.}
     \label{fig:Chicago}
 \end{figure}

 Figure~\ref{fig:Chicago} presents the reconstructed transmission parameter considering the values of the regularization parameters $\alpha=10^{-1}$, $10^{-2}$, $10^{-3}$, $10^{-4}$, and $10^{-5}$. It also presents the corresponding model in-sample predictions of cases compared with the observed daily number of infections. For regularization parameter values smaller or equal to $10^{-2}$, the model predictions were adherent to the observed data. The large values at the beginning of the series of reconstructed transmissions are outliers, caused, basically, by poor data collection. This pattern was observed using different models and estimation techniques \cite{albani2022c}. The estimated parameter was generally smooth, using different values for the regularization parameter.

 \paragraph*{Canada in 2021}
Now, $\beta$ is estimated from infections in Canada, during 2021. The dataset is composed by the daily infections, from 01-Jan-2021 to 31-Dec-2020. Vaccination started in 15-Jun-2021, with a time varying rate. The time series were also smoothed by a 7-day moving average, and the step size used in estimation was of ten days. Again, to solve the model with a step size of one day, the transmission parameter was linearly interpolated.

 \begin{figure}[!htb]
     \centering
     \includegraphics[width=0.45\textwidth]{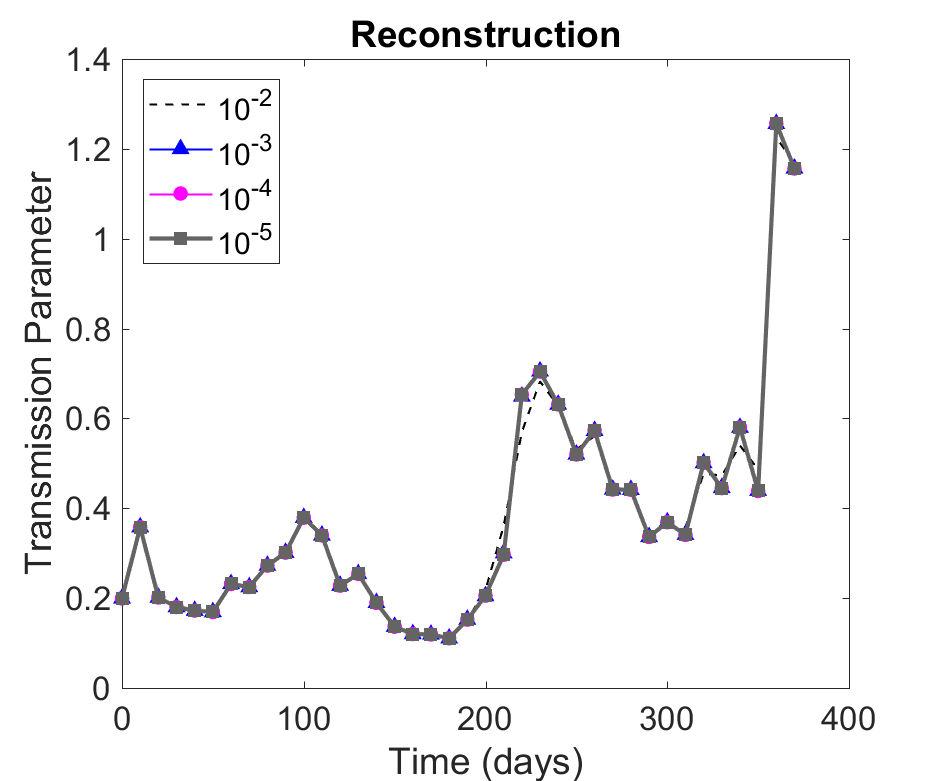}\hfill
     \includegraphics[width=0.45\textwidth]{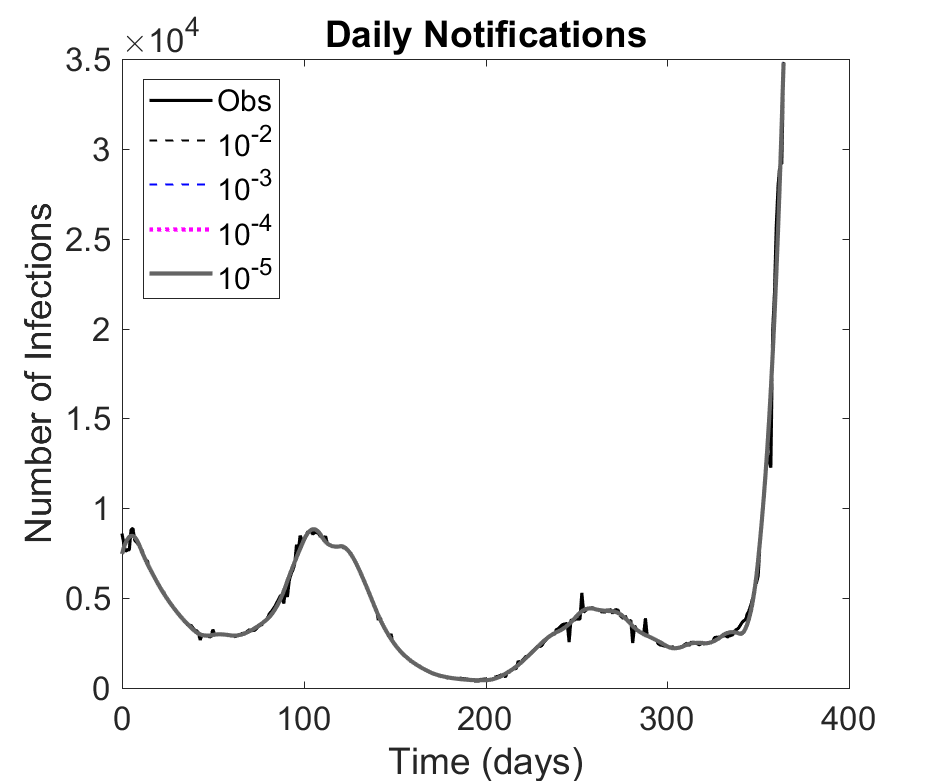}\hfill
     \caption{Left: Reconstructed transmission parameters for different regularization parameter values. Right: Comparison of the corresponding model predictions of cases with the observed daily infections. COVID-19 infections in Canada, during 2021.}
     \label{fig:Canada}
 \end{figure}

The reconstructed parameters obtained using the values of the regularization parameter $10^{-2}$, $10^{-3}$, $10^{-4}$, and $10^{-5}$ can be seen in Fig.~\ref{fig:Chicago}. In the same figure, the corresponding model in-sample predicted infections and the observed daily notifications can also be observed. The model predictions were adherent to the observed data. To account for vaccination, the parameter $\nu$ was set equal to the daily number of vaccinated divided by the population of Canada in 2021, which was approximately 38 million. This leads to approximately 17 million susceptible individuals immunized with the vaccine, which is about 72\% of the total number of vaccinated. In principle, it considers that vaccines are not 100\% effective and individual already immunized by recovering from an undetected infection can be vaccinated. In this example, again, the estimated transmission was smooth, using different values of the regularization parameters.

\section{Conclusions}
The article proposed an SEIR-like epidemiological analysis to describe the spread of an infectious disease, such as COVID-19, accounting for asymptomatic infections, vaccination, death, and different levels of disease severity. The model parameters were continuous functions. The regularity of the parameter-to-solution map that associates the transmission parameter with the model solution was extensively analyzed. Sufficient conditions for continuity and Fr\'echet differentiability were stated. The corresponding inverse problem was also analyzed. Tikhonov-type regularization was applied, leading to the existence, stability, and convergence of regularized solutions. The model estimation was tested with synthetic and real data. For synthetic data, the transmission parameter was successfully estimated from noisy data and perturbed parameters. The transmission parameter was also successfully estimated from the reported infections in Chicago, during 2020, and Canada, during 2021, when a massive vaccination strategy was implemented. In both cases, the model in-sample predictions were adherent to the dataset and the estimated transmissions varied smoothly with time.

\bibliographystyle{amsplain}
\bibliography{sample}

\providecommand{\bysame}{\leavevmode\hbox to3em{\hrulefill}\thinspace}
\providecommand{\MR}{\relax\ifhmode\unskip\space\fi MR }
\providecommand{\MRhref}[2]{%
  \href{http://www.ams.org/mathscinet-getitem?mr=#1}{#2}
}
\providecommand{\href}[2]{#2}
\begin{thebibliography}{10}

\bibitem{abate2020}
S.~Abate, S.~Ahmed~Ali, B.~Mantfardo, and B.~Basu, \emph{Rate of intensive care
  unit admission and outcomes among patients with coronavirus: A systematic
  review and meta-analysis}, PloS One \textbf{15} (2020), no.~7, e0235653.

\bibitem{achterberg2020}
M.~Achterberg, B.~Prasse, L.~Ma, S.~Trajanovski, M.~Kitsak, and P.~Van~Mieghem,
  \emph{{Comparing the accuracy of several network-based COVID-19 prediction
  algorithms}}, International Journal of Forecasting (2020).

\bibitem{albani2022role}
V.~Albani, R.~Albani, N.~Bobko, E.~Massad, and J.~Zubelli, \emph{{On the role
  of financial support programs in mitigating the SARS-CoV-2 spread in
  Brazil}}, BMC Public Health \textbf{22} (2022), no.~1, 1--17.

\bibitem{albani2022}
V.~Albani, R.~Albani, E.~Massad, and J.~Zubelli, \emph{{Nowcasting and
  Forecasting COVID-19 Waves: The Recursive and Stochastic Nature of
  Transmission}}, Royal Society Open Science \textbf{9} (2022), 220489.

\bibitem{albani2022c}
V.~Albani, M.~Grasselli, W.~Peng, and J.~Zubelli, \emph{{The Interplay between
  COVID-19 and the Economy in Canada}}, Journal of Risk and Financial
  Management \textbf{15} (2022), no.~10.

\bibitem{albani2021covid}
V.~Albani, J.~Loria, E.~Massad, and J.~Zubelli, \emph{{COVID}-19 underreporting
  and its impact on vaccination strategies}, BMC infectious diseases
  \textbf{21} (2021), no.~1, 1--13.

\bibitem{Albani2021}
\bysame, \emph{The impact of {COVID}-19 vaccination delay: A data-driven
  modeling analysis for {C}hicago and {N}ew {Y}ork {C}ity}, Vaccine \textbf{39}
  (2021), no.~41, 6088--6094.

\bibitem{Albani2020}
V.~Albani, R.~Velho, and J.~Zubelli, \emph{Estimating, monitoring, and
  forecasting {COVID}-19 epidemics: a spatiotemporal approach applied to {NYC}
  data}, Scientific Reports \textbf{11} (2021), no.~1, 1--15.

\bibitem{Albani2024}
Vinicius~VL Albani and Jorge~P Zubelli, \emph{Stochastic transmission in
  epidemiological models}, Journal of Mathematical Biology \textbf{88} (2024),
  no.~3, 25.

\bibitem{beretta2021}
A.~Aspri, E.~Beretta, A.~Gandolfi, and E.~Wasmer, \emph{{Mortality containment
  vs. economics opening: optimal policies in a SEIARD model}}, Journal of
  Mathematical Economics \textbf{93} (2021), 102490.

\bibitem{athayde2022}
G.~Athayde and A.~Alencar, \emph{{Forecasting Covid-19 in the United Kingdom: A
  dynamic SIRD model}}, PLoS ONE \textbf{17} (2022), e0271577.

\bibitem{bachman}
George Bachman and Lawrence Narici, \emph{Functional {A}nalysis}, Academic
  Press, 1966.

\bibitem{B1}
N.~Bellomo, R.~Bingham, M.~Chaplain, G.~Dosi, G.~Forni, D.~Knopoff,
  J.~Lowengrub, R.~Twarock, and M.~Virgillito, \emph{A multiscale model of
  virus pandemic: {H}eterogeneous interactive entities in a globally connected
  world}, Mathematical Models and Methods in Applied Sciences \textbf{30}
  (2020), no.~08, 1591--1651.

\bibitem{bellomo2}
N.~Bellomo, D.~Burini, and N.~Outada, \emph{{Multiscale models of Covid-19 with
  mutations and variants}}, Networks and Heterogeneous Media \textbf{17}
  (2022), no.~3, 293.

\bibitem{bertozzi2020}
A.~Bertozzi, E.~Franco, G.~Mohler, M.~Short, and D.~Sledge, \emph{{The
  challenges of modeling and forecasting the spread of COVID-19}}, Proceedings
  of the National Academy of Sciences \textbf{117} (2020), no.~29,
  16732--16738.

\bibitem{brezis2011}
Haim Brezis, \emph{Functional analysis, {S}obolev spaces and partial
  differential equations}, vol.~2, Springer, 2011.

\bibitem{byambasuren2020}
Oyungerel Byambasuren, Magnolia Cardona, Katy Bell, Justin Clark, Mary-Louise
  McLaws, and Paul Glasziou, \emph{Estimating the extent of asymptomatic
  {COVID}-19 and its potential for community transmission: systematic review
  and meta-analysis}, Official Journal of the Association of Medical
  Microbiology and Infectious Disease Canada \textbf{5} (2020), no.~4,
  223--234.

\bibitem{somersalo2020}
D~Calvetti, A~Hoover, J~Rose, and E~Somersalo, \emph{Bayesian particle filter
  algorithm for learning epidemic dynamics}, Inverse Problems \textbf{37}
  (2021), no.~11, 115008.

\bibitem{calvetti2020}
Daniela Calvetti, Alexander~P Hoover, Johnie Rose, and Erkki Somersalo,
  \emph{Metapopulation network models for understanding, predicting, and
  managing the coronavirus disease {COVID}-19}, Frontiers in Physics \textbf{8}
  (2020), 261.

\bibitem{campos2021}
Eduardo~Lima Campos, Rubens~Penha Cysne, Alexandre~L Madureira, and
  G{\'e}lcio~LQ Mendes, \emph{Multi-generational {SIR} modeling:
  {D}etermination of parameters, epidemiological forecasting and age-dependent
  vaccination policies}, Infectious Disease Modelling \textbf{6} (2021),
  751--765.

\bibitem{cartan1971}
Henri Cartan, \emph{Differential {C}alculus. {H}ermann}, Houghton Mifflin Co.,
  Paris/Boston, MA, 1971.

\bibitem{gatto2020}
M.~Gatto, E.~Bertuzzo, L.~Mari, S.~Miccoli, L.~Carraro, R.~Casagrandi, and
  A.~Rinaldo, \emph{Spread and dynamics of the covid-19 epidemic in italy:
  Effects of emergency containment measures}, Proceedings of the National
  Academy of Sciences \textbf{117} (2020), no.~19, 10484--10491.

\bibitem{grasselli2020}
Giacomo Grasselli, Alberto Zangrillo, Alberto Zanella, Massimo Antonelli, Luca
  Cabrini, Antonio Castelli, Danilo Cereda, Antonio Coluccello, Giuseppe Foti,
  Roberto Fumagalli, et~al., \emph{Baseline characteristics and outcomes of
  1591 patients infected with {SARS}-{C}o{V}-2 admitted to {ICU}s of the
  {L}ombardy {R}egion, {I}taly}, Jama \textbf{323} (2020), no.~16, 1574--1581.

\bibitem{guan2020}
W.~Guan, Z.~Ni, Y.~Hu, W.~Liang, C.~Ou, J.~He, L.~Liu, H.~Shan, C.~Lei, D.~Hui,
  et~al., \emph{{Clinical characteristics of coronavirus disease 2019 in
  China}}, New England Journal of Medicine \textbf{382} (2020), no.~18,
  1708--1720.

\bibitem{guglielmi2022}
N.~Guglielmi, E.~Iacomini, and A.~Viguerie, \emph{{Delay differential equations
  for the spatially resolved simulation of epidemics with specific application
  to COVID-19}}, Mathematical Methods in the Applied Sciences (2022).

\bibitem{huang2020}
Chaolin Huang, Yeming Wang, Xingwang Li, Lili Ren, Jianping Zhao, Yi~Hu,
  Li~Zhang, Guohui Fan, Jiuyang Xu, Xiaoying Gu, et~al., \emph{Clinical
  features of patients infected with 2019 novel coronavirus in {W}uhan,
  {C}hina}, The lancet \textbf{395} (2020), no.~10223, 497--506.

\bibitem{keeling2008}
MJ~Keeling and P~Rohani, \emph{Modeling {I}nfectious {D}iseases in {H}umans and
  {A}nimals}, Princeton, NJ (2008).

\bibitem{kerr2021}
C.~Kerr, R.~Stuart, D.~Mistry, R.~Abeysuriya, K.~Rosenfeld, G.~Hart,
  R.~N{\'u}{\~n}ez, J.~Cohen, P.~Selvaraj, B.~Hagedorn, et~al., \emph{{Covasim:
  an agent-based model of COVID-19 dynamics and interventions}}, PLOS
  Computational Biology \textbf{17} (2021), no.~7, e1009149.

\bibitem{lauer2020}
Stephen~A Lauer, Kyra~H Grantz, Qifang Bi, Forrest~K Jones, Qulu Zheng,
  Hannah~R Meredith, Andrew~S Azman, Nicholas~G Reich, and Justin Lessler,
  \emph{The incubation period of coronavirus disease 2019 ({COVID}-19) from
  publicly reported confirmed cases: estimation and application}, Annals of
  internal medicine \textbf{172} (2020), no.~9, 577--582.

\bibitem{namasudra2021}
S.~Namasudra, S.~Dhamodharavadhani, and R.~Rathipriya, \emph{{Nonlinear neural
  network based forecasting model for predicting COVID-19 cases}}, Neural
  Processing Letters (2021), 1--21.

\bibitem{scherzer2009}
Otmar Scherzer, Markus Grasmair, Harald Grossauer, Markus Haltmeier, and Frank
  Lenzen, \emph{Variational methods in imaging}, Springer, 2009.

\bibitem{sotomayor2011}
Jorge Sotomayor, \emph{Equa{\c{c}}oes diferenciais ordin{\'a}rias}, Sao Paulo:
  Editora Livraria da F{\i}sica, 2011.

\bibitem{stewart2022}
R.~Stewart, S.~Erwin, J.~Piburn, N.~Nagle, J.~Kaufman, A.~Peluso, J.~Christian,
  J.~Grant, A.~Sorokine, and B.~Bhaduri, \emph{{Near real time monitoring and
  forecasting for COVID-19 situational awareness}}, Applied Geography
  \textbf{146} (2022), 102759.

\bibitem{teschl2012}
Gerald Teschl, \emph{Ordinary differential equations and dynamical systems},
  vol. 140, American Mathematical Soc., 2012.

\bibitem{WHO}
WHO, \emph{Coronavirus disease ({COVID-19}) pandemic},
  \url{https://www.who.int/europe/emergencies/situations/covid-19}, Accessed:
  2024-08-14.

\bibitem{who19}
\bysame, \emph{Report of the {WHO}-{C}hina joint mission on coronavirus disease
  2019 ({COVID}-19)}, 2020.

\bibitem{worldometer}
Worldometer, \emph{{COVID-19 Coronavirus Pandemic}},
  \url{https://www.worldometers.info/coronavirus/}, Accessed: 2024-08-14.

\end{thebibliography}

\end{document}